\documentclass{article}

\usepackage[a4paper, total={5.5in, 8.15in}]{geometry}
\usepackage[utf8]{inputenc} 
\usepackage[T1]{fontenc}    
\usepackage{url}            
\usepackage{booktabs}       
\usepackage{amsfonts}       
\usepackage{nicefrac}       
\usepackage{microtype}      
\usepackage{lipsum}
\usepackage{float}
\usepackage{caption}
\usepackage{amsmath}
\usepackage{lscape}
\usepackage{setspace}
\usepackage{natbib}

\usepackage{eurosym}
\usepackage[pdftex]{graphicx}
\usepackage[T1]{fontenc}
\usepackage[utf8]{inputenc}
\usepackage{authblk}
\usepackage{dcolumn}
\usepackage{multirow}
\usepackage{hyperref}
\hypersetup{colorlinks=true,linkcolor=blue,urlcolor=blue,anchorcolor=blue,citecolor=blue}
\newcommand{\sym}[1]{{#1}} 
\usepackage{todonotes}
\providecommand{\keywords}[1]
{
  \small	
  \textbf{\textit{Keywords---}} #1
}

\title{The Economics and Econometrics of \\ Gene--Environment Interplay}

\author{
 Pietro Biroli\thanks{University of Bologna. E-mail: \href{mailto:Pietro.Biroli@econ.uzh.ch}{pietro.biroli@unibo.it}.}, \hspace{0.5cm} Titus Galama\thanks{University of Southern California and VU University Amsterdam. E-mail: \href{mailto:galama@usc.edu}{galama@usc.edu}}, \hspace{.5cm} Stephanie von Hinke\thanks{University of Bristol; Erasmus University Rotterdam; Institute for Fiscal Studies. E-mail: \href{mailto:S.vonHinke@bristol.ac.uk}{S.vonHinke@bristol.ac.uk}.}, \\ \hspace{.5cm} Hans van Kippersluis\thanks{Corresponding author: Erasmus University Rotterdam. E-mail:  \href{mailto:hvankippersluis@ese.eur.nl}{hvankippersluis@ese.eur.nl}.}, \hspace{.5cm} Cornelius A. Rietveld\thanks{Erasmus University Rotterdam. E-mail:  \href{mailto:nrietveld@ese.eur.nl}{nrietveld@ese.eur.nl}.}, \hspace{.5cm} Kevin Thom\thanks{University of Wisconsin. E-mail: \href{mailto:thomk@uwm.edu}{thomk@uwm.edu}}

}

\date{\today \thanks{Acknowledgements: We thank the \href{https://gene-environment.com/}{GEIGHEI} members for many discussions on $G \times E$ interplay and, in particular, Rita Dias Pereira for creating the polygenic indices used in this study. We gratefully acknowledge financial support from NORFACE DIAL (462-16-100). Research reported in this publication was also supported by the European Research Council (DONNI 851725 and GEPSI 946647), the National Institute on Aging of the National Institutes of Health (RF1055654 and R56AG058726), and the Dutch National Science Foundation (016.VIDI.185.044). The UK Medical Research Council and Wellcome (Grant ref: 217065/Z/19/Z) and the University of Bristol provide core support for ALSPAC. This publication is the work of the authors who will serve as guarantors for the contents of this paper. A comprehensive list of grants funding is available on the ALSPAC website (http://www.bristol.ac.uk/alspac/external/documents/grant-acknowledgements.pdf). GWAS data was generated by Sample Logistics and Genotyping Facilities at Wellcome Sanger Institute and LabCorp (Laboratory Corporation of America) using support from 23andMe. Consent for biological samples has been collected in accordance with the Human Tissue Act (2004). We are extremely grateful to all the families who took part in this study, the midwives for their help in recruiting them, and the whole ALSPAC team, which includes interviewers, computer and laboratory technicians, clerical workers, research scientists, volunteers, managers, receptionists and nurses. Ethical approval for the study was obtained from the ALSPAC Ethics and Law Committee and the Local Research Ethics Committees. Informed consent for the use of data collected via questionnaires and clinics was obtained from participants following the recommendations of the ALSPAC Ethics and Law Committee at the time.}}

\begin{document}
\maketitle

\begin{abstract}
\noindent Economists and social scientists have debated the relative importance of nature (one's genes) and nurture (one's environment) for decades, if not centuries. This debate can now be informed by the ready availability of genetic data in a growing number of social science datasets. This paper explores the potential uses of genetic data in economics, with a focus on estimating the interplay between nature (genes) and nurture (environment). We discuss how economists can benefit from incorporating genetic data into their analyses even when they do not have a direct interest in estimating genetic effects. We argue that gene--environment ($G \times E$) studies can be instrumental for (i) testing economic theory, (ii) uncovering economic or behavioral mechanisms, and (iii) analyzing treatment effect heterogeneity, thereby improving the understanding of how (policy) interventions affect population subgroups. We introduce the reader to essential genetic terminology, develop a conceptual economic model to interpret gene--environment interplay, and provide practical guidance to empirical researchers. 
\end{abstract}

\keywords{Gene-by-Environment Interplay; \and Polygenic Indices; \and Social Science Genetics \and ALSPAC} \\
\indent \textbf{\textit{JEL Classifications}: D1, D3, I1, I2, J1} 

\bigskip 

\clearpage
\onehalfspacing

\section{Introduction}
\label{sec:intro}

The debate on the relative importance of nature versus nurture in the development of human traits is amongst the oldest in the social sciences. Decades' worth of studies on twins provide evidence that genetic factors are responsible for significant variation in outcomes of interest to economists, including educational attainment, smoking, obesity, risk-taking, income, wealth, health, and many more.  Approximately 25-75\% of the variation in a wide range of key behaviors, traits, and outcomes can be attributed to genetic differences \citep{Polderman2015}.  In fact, a large body of evidence stemming from heritability studies has been synthesized into the first law of behavioral genetics stating that ``\textit{all human behavioural traits are heritable}'' \citep{Turkheimer2000}. Furthermore, researchers increasingly recognize that the traditional notion that nature and nurture operate independently is obsolete \citep{Plomin1977,Turkheimer2000,Rutter2006}. Pitting nature \textit{against} nurture should be relinquished in favor of a view that considers a more complex interplay that may exist between people's genetic makeup and the environment in which they develop \citep{Hunter2005, Heckman2007}. 

Economic modeling of a more complex gene--environment ($G \times E$ [pronounced ``G-by-E'']) interplay is now feasible thanks to recent advances that have significantly reduced the barriers to incorporating genetic data into economic analyses. As the cost of measuring genetic variation across people continues to fall, there has been rapid growth in the availability of human molecular genetic data. These data allow researchers to include in their analysis specific genetic variants (so-called single-nucleotide polymorphisms or SNPs [pronounced ``snips'']) as well as indices of, typically, very large numbers of genetic variants, called  polygenic scores (PGSs) or polygenic indices (PGIs). These PGIs have substantially greater predictive power than single genetic variants, and they are currently readily available to users in a number of rich longitudinal datasets of particular relevance to economists: the Health and Retirement Study (HRS), the National Longitudinal Study of Adolescent to Adult Health (AddHealth), the Panel Study of Income Dynamics (PSID), and the English Longitudinal Study of Aging (ELSA) \citep{Becker2021}, among many others. As a result, economists now have at their disposal a growing number of datasets containing genetic measures that can significantly predict a range of behaviors and outcomes. For example, the PGI for educational attainment currently explains around 12-16\% of the variation in educational attainment, which is on par with some of the strongest environmental determinants such as parental education and income \citep{lee2018gene, Okbay2022}. Since anyone can now explore the importance of gene--environment interplay, even without much knowledge of genetics, it is important for economists to understand the complexities that come with the analysis of genetic data and the interpretation of results. This essential guidance is what we aim to provide here.

This article considers the economics and econometrics of $G \times E$ interplay. Such interplay encompasses the main effects of $G$ and $E$ as well as their potential interaction. $G \times E$ interaction occurs when environmental factors influence the relationship between genetic factors and a particular outcome of interest, or vice versa.\footnote{Gene--environment interplay differs from the study of epigenetics, which focuses on the role of gene expression. Whereas individuals' genetic makeup is fixed at conception, their gene expression may change over the life course due to environmental influences. As such, epigenetic processes can provide one explanation for the existence of gene--environment interplay, but they are not the only possible mechanism. A detailed discussion of epigenetics is beyond the scope of this paper.
} 
While uncovering $G \times E$ interaction effects is of interest, it is not the only focus of $G \times E$ interplay studies, as main effects are relevant too. Our study builds upon and extends general surveys on the promise of using genetic data in economic analyses (such as those found in \citet{Benjamin2011} and \citet{Beauchamp2011} 
and more specifically the surveys on $G \times E$ interplay (as in \citet{Fletcher2013} and \citet{Schmitz2017}).\footnote{See \cite{Rutter2006}, \citet{Plomin2014}, \citet{Mills2020book} and \citet{domingue2020interactions} for related work in other disciplines.}

The aim of this article is twofold: first, to introduce the reader to key concepts and recent developments in the field of $G \times E$ interplay and, second, to offer practical guidance to empirical researchers interested in using available genetic data to explore the nature--nurture interplay in behaviors and outcomes. As part of our analysis, we develop an economic model of decision-making with genetic heterogeneity to demonstrate how economic theory can guide empirical $G \times E$ analyses and help in the interpretation of $G \times E$ findings. We believe an improved comprehension of $G \times E$ interplay represents a basic scientific advance in our understanding of the role of nature and nurture in shaping human capabilities. This in turn may lead to unforeseen advances in scientific knowledge and to novel applications. However, even without an inherent interest in the underlying biology of economic behavior, we contend that empirical work on $G \times E$ interplay is of general interest to economists for at least three reasons:

(1) \textit{Testing theoretical predictions}:  Economic theories often predict that individuals will differ in their response to a common environmental change because of idiosyncratic characteristics like preferences, health endowments, abilities, etc.  For example, many economic theories of human capital production assume that acquired abilities and endowments raise the productivity of later investments \citep[e.g.,][]{ben1967production,becker1986human,cunha2007technology}, and that parents will respond to the endowments of their children \citep[e.g.,][]{becker1976child,behrman1997intrahousehold,CURRIE20111315}. Since primitive variables such as endowments and abilities are typically hard to measure, testing such predictions can be challenging.  For example, in the child development literature,  childhood endowments are often proxied by conditions such as birth weight or neonatal health shocks.  Although such proxies are undoubtedly valuable, they tend to capture a limited set of acute conditions and are rarely independent of (prenatal) parental investments.
Observable genetic variation offers a new and powerful way to measure some of these previously unobserved characteristics.  The presence (or absence) of $G \times E$ interplay can thus provide clarifying evidence on theoretical predictions. For example, \citet{Muslimova2020b} find empirical support for complementarity in skill formation by exploiting random between-sibling variation in genetic endowments and environments, and \citet{breinholt2020child}, \citet{Sanz-de-galdeano2019}, \citet{houmark2020nurture} and \citet{fletcher2020production} provide evidence that parental investments respond to children's genetic endowments.

(2) \textit{Uncovering mechanisms}: Evidence on $G \times E$ interplay can also provide clues with respect to the economic or behavioral mechanisms through which genetic factors operate. For example, \citet{barth2020genetic} find evidence that access to defined benefit pension plans substantially moderates the relationship between a measure of the genetic propensity to stay in school longer and household wealth. This finding suggests that genetic endowments may operate through mechanisms that govern financial decision-making and portfolio choice. Learning more about the mechanisms associated with specific genetic factors can in turn help economists decide on the most realistic way to incorporate individual-level heterogeneity into, e.g., structural estimations of life-cycle models \citep{Benjamin2012}.

(3) \textit{Assessing treatment effect heterogeneity}: Although there has been heated debate about the policy relevance of heritability studies \citep{Goldberger1979,taubman1981heritability,Manski2011}, understanding more about $G \times E$ interplay can provide novel evidence of value to policy-makers.  The study of $G \times E$ interplay may help identify environments or policies that reduce genetic disadvantage \citep{Barcellos2018,Barcellos2021} or characterize the kinds of individuals who thrive under different sets of political and economic institutions \citep{Rimfeld2018}.  The study of treatment effect heterogeneity stemming from genetic factors is particularly relevant to studies of intergenerational mobility. For example, if a certain policy fails to have its intended effect on a target group with a specific genetic predisposition, then this may propagate across generations (as genes are passed on to offspring), potentially explaining increasing intergenerational inequalities.

Evidence on $G \times E$ interplay may in the future contribute to the development of personalized interventions tailored to individual characteristics \citep{Benjamin2012}. 
Enthusiasm over this use of molecular genetic results is still largely premature since the current predictive power of genetic measures precludes accurate \textit{individual}-level prediction of future traits such as disease or economic outcomes \citep{morris2020can,Turley2021embryo}.  However, these measures perform well in detecting \textit{population}-level average relationships.  While molecular genetic measures may currently have limited value for targeting interventions to specific individuals, they have proven to be useful for our understanding of the distributional consequences of economy-wide policies.

Finally, just as the analysis of $G \times E$ interplay can provide novel insights for economists, there is great potential to using the toolbox of economics to better design $G \times E$ studies and thus to advance the fields of genetics in general and social-science genetics in particular. Since $G \times E$ interplay often stems from endogenous behavioral adjustments, economic theory can help clarify why and when such interplay might occur and what it implies for policy. Empirically, both genetic endowments and environmental factors are typically endogenous in the study of a particular outcome. Ongoing advances in methods and data are making possible causal inferences of genetic factors. Economists have substantial experience with exploiting exogenous variation in environmental exposures and have developed a large toolbox to deal with endogeneity. Given the importance of establishing which causal environmental exposures moderate genetic predispositions, economists are well positioned to improve understanding of the complex interplay between nature and nurture in shaping life outcomes. 

Besides providing an introduction to and overview of the current state of the field and the various new directions it is taking, this review also makes several novel contributions. In \hyperref[sec:econModel]{Section~\ref*{sec:econModel}}, we present a stylized economic model of a behavioral choice (e.g., an investment in education subject to a budget constraint) that produces $G \times E$ interplay. This stylized economic theory highlights the role of behavioral choices in response to genes and environments in generating $G \times E$ interplay. In \hyperref[sec:specification]{Section~\ref*{sec:specification}}, we discuss the intricacies of interpreting an empirical model that seeks to estimate $G \times E$, providing a systematic categorization of the various types of $G \times E$ analyses and a discussion of the direction and nature of bias in $G$, $E$ and $G \times E$ with respect to the ideal (unbiased) case in which both $G$ and $E$ are exogenous. Finally, in \hyperref[sec:application]{Section~\ref*{sec:application}} we provide an illustration of a  $G \times E$ estimation, and uncover a novel $G \times E$ interaction between being old for grade in school ($E$; exogenous due to sharp cut offs in month of birth determining earliest eligibility for school entry) and the genetic propensity for educational attainment ($G$) on test scores at different ages throughout childhood. All syntax for the empirical analysis is included in our \href{http://github.com/geighei/GxE_4practitioners}{GitHub} repository, and we discuss the measurement of $G$, with a main focus on PGIs, in \hyperref[sec:G]{Section~\ref*{sec:G}} and \hyperref[sec:glossary]{Appendix~\ref*{sec:glossary}}.

\section{An economic model of \texorpdfstring{$G \times E$}{} interplay} \label{sec:econModel}
In this section, we introduce a stylized economic model of a behavioral choice with genetic and environmental factors to elucidate the different ways in which $G \times E$ interplay may manifest. Let $Y_{i}$ represent an outcome of interest, such as educational attainment.  We assume that $Y_{i}$ is produced as a function of environmental factors $E_{i}$, genetic factors $G_{i}$, a choice or investment that individuals make $x_{i}$, and a vector of shocks or random components $e_{i}$.  Let $F(x_{i},G_{i},E_{i},e_{i})$ represent the production function for $Y_{i}$. In choosing $x_i$, individuals maximize utility, which we assume here is simply the difference between $Y_{i}=F(x_{i},G_{i},E_{i},e_{i})$ and a cost function $C(x_i,G_i,E_i,e_{i})$.  Costs here can be interpreted broadly as including monetary costs and time and effort. To fix ideas, let us suppose that $Y_{i}$ represents years of schooling, $E_{i}$ measures the quality (environment) of schools available to individual $i$, $G_i$ denotes the genetic factors that are predictive of years of schooling, and $x_{i}$ represents academic effort (e.g., time and effort spent studying). 

The individual's decision problem can then be represented as
\begin{equation}
\max_{x_{i}}\:\:F(x_{i},G_{i},E_{i},e_{i})-C(x_{i},G_{i},E_{i},e_{i}).
\label{eq:valuefunction}
\end{equation}
Even this simple setup highlights the complexity of behavioral responses to genetic endowments and environments. Genetic factors can influence an individual's efficiency in producing $Y_{i}$ or an individual's preferences for engaging in activities that produce $Y_{i}$, a point made by \citet{Biroli2015} in the context of obesity. The interplay between genes and environments can therefore arise within the production function (capturing efficiency), within the cost function (capturing preferences), and through interactions between the two that arise as the individual endogenously chooses $x_{i}$.\footnote{Choosing how much to invest in academic effort $x_i$ in practice is an intertemporal maximization problem, with the costs being mostly immediate and the benefits reaped in the future. In reality, therefore, time and risk preferences also play a major role, and these are in turn known to be partially driven by genetic and environmental factors. For simplicity, we ignore this dynamic element here.} The first-order condition for an optimum requires
\begin{equation}
    F_{x}(x_{i},G_{i},E_{i},e_{i})-C_{x}(x_{i},G_{i},E_{i},e_{i})=0,
    \label{eq:FOC}
\end{equation}
while the second-order condition requires
\begin{equation}
    F_{xx}(x_{i},G_{i},E_{i},e_{i})-C_{xx}(x_{i},G_{i},E_{i},e_{i})<0.
    \label{eq:SOC}
\end{equation}
The solution to the first-order condition defines optimal investment $x_i^{*}$ as a function of genes and environments or $x_i^{*}=x_i^{*}(G_{i},E_{i},e_{i})$. The second-order condition guarantees that the solution is unique (if it holds for all $x_i$). 

\paragraph{Reinforcement or substitution:} To understand whether behavioral responses $x_i^*$ reinforce or substitute the effect of genetic and environmental factors, we first consider the effects of environmental factors $E_i$ and genetic factors $G_i$ separately as
\begin{eqnarray}
\label{eq:FE}
\frac{\partial Y_i}{\partial E_i} = F_E + F_x\,x^*_{i,E}, \\
\label{eq:FG}
\frac{\partial Y_i}{\partial G_i} = F_G + F_x\,x^*_{i,G}. 
    \end{eqnarray}
The first term on the right-hand side of \autoref{eq:FE} and \autoref{eq:FG} represents the effect of better environmental ($E_i$) or genetic ($G_i$) factors, with investment (effort, $x_i^*$) held constant. Without loss of generality, we assume that $E_{i}$ and $G_{i}$ are measured such that $F_E(\cdot)$ and $F_G(\cdot)$ are positive. In other words, higher quality schools and certain genetic factors make learning more efficient. Given that effort is costly, whether an individual increases her effort in response to better environments ($x_{i,E}^*(\cdot)=\left(\partial x_i^*/\partial E\right)>0$) or better genetic endowments ($x_{i,G}^*(\cdot)=\left(\partial x_i^*/\partial G\right)>0$) depends on how marginal benefits and marginal costs vary with effort and with environmental and genetic factors. Thus, behavioral responses $x_i^*$ can be compensatory or reinforcing with respect to both genetic endowments and the quality of the environments.

Differentiating \autoref{eq:FOC} with respect to $E_i$ (recognizing that $x_{i}^{*}$ is a function of $E_i$) yields the following expression for the optimal response of $x_{i}^*$ to a ceteris paribus change in the environment $E_i$:
\begin{equation}
    x_{i,E}^{*}=\frac{C_{xE}-F_{xE}}{F_{xx}-C_{xx}}.  \label{eq:xE}
\end{equation}
Similarly, one can differentiate \autoref{eq:FOC} with respect to $G_i$ to arrive at the expression for the optimal behavioral response to a ceteris paribus change in genetic factor $G_i$:
\begin{equation}
    x_{i,G}^{*}=\frac{C_{xG}-F_{xG}}{F_{xx}-C_{xx}} \label{eq:xG}.
\end{equation}

Given the second-order condition (\autoref{eq:SOC}), we have $(F_{xx}-C_{xx})<0$. Hence, the signs of the partial derivatives in \autoref{eq:xE} and \autoref{eq:xG} depend on the signs of $(C_{xE}-F_{xE})$ and $(C_{xG}-F_{xG})$, respectively. For example, suppose that $E_i$ measures the quality of schools, which would reduce the disutility of school effort ($C_{xE}<0$) and increase the marginal productivity of effort in school ($F_{xE}>0)$.  Then, an improvement in the environment ($E_i > 0$) would reinforce changes in behavior: $x_i>0$.  On the other hand, when $E$ captures something like the strictness of school discipline, the marginal disutility of effort increases ($C_{xE}>0$) but the marginal productivity of effort investments may also increase ($F_{xE}>0$). In this case, the effect of an increase in $E_i$ is ambiguous as there are offsetting effects, allowing for both reinforcement and substitution.   

\paragraph{Gene--environment interplay:} Typically, gene-by-environment interaction is said to be present when the relationship between $G_{i}$ and $Y_{i}$ is affected by the level of $E_{i}$---or vice versa, when the relationship between $E_{i}$ and $Y_{i}$ is affected by the level of $G_{i}$.  We can formalize this as a statement about the expected value of the following cross-partial derivative:
\begin{equation}
\frac{\partial^{2} Y_{i} }{\partial G_{i} \partial E_{i}} = \underbrace{F_{GE}}_{\text{Tech $G \times E$}}+\underbrace{F_{Gx}\,x^*_{i,E}}_{\text{G-Choice Comp}}+\underbrace{F_{Ex}\,x^*_{i,G}}_{\text{E-Choice Comp}}+\underbrace{F_{x}x^*_{i,GE}}_{\text{Choice $G \times E$}}+\underbrace{F_{xx}\,x^*_{i,E} \,x^*_{i,G}}_{\text{Tech. Nonlinearities}}. \label{eq:GEDecomp}
\end{equation}
Each term in this expression represents a distinct mechanism through which genes and environments can interact in the presence of endogenous choices. The first term on the right-hand side represents \textit{technological G $\times$ E}, or interactions that occur between genes and environments at the level of the production function $F(\cdot)$, with choices $x_i$ held constant. In our example, this could arise if the instruction at higher quality schools $E_i$ benefits children with higher (or lower) levels of $G_{i}$, when child study effort $x_i$ is held fixed. Better quality schools might be able to offer lessons and teacher interactions that are more productive for everyone, but they might particularly help children with higher or lower genetic endowments $G_{i}$.  

The other four terms in \autoref{eq:GEDecomp} all represent gene--environment interactions mediated by responses in optimal behavior. The second term represents \textit{gene--choice complementarity}.  Such an interaction arises when changes in the environment $E_i$ induce changes in the choice $x_{i}$ and there exists a complementarity between individual choice $x_i$ and genetic endowments $G_i$ in the production function. In our example, this case might arise if better schools induce children to exert more effort ($x^*_{i,E}>0$) and if extra effort is more productive in building human capital for individuals with higher genetic endowments ($F_{Gx}>0$). Similarly, the third term represents \textit{environment--choice complementarity}. Such an interaction arises if, for example, higher levels of $G_i$ induce individuals to choose higher levels of $x_i$ and if higher levels of $E$ complement greater individual investment in the production function. In our example, this would occur if $G_i$ operates by making it easier for individuals to supply effort $x_i$ (e.g., genes associated with improved focus or determination) and if better quality schools $E_i$ particularly reward this effort. 

The fourth term in \autoref{eq:GEDecomp} represents an interaction generated by a complementarity between genes and environment at the level of individual optimal choices, or \textit{Choice G $\times$ E}. For example, it could be the case that higher levels of $G_i$ reduce the costs of supplying effort (increasing $x_i$, ceteris paribus) or that better schools $E_i$  work by encouraging students to supply more effort (also increasing $x_i$, ceteris paribus).  In the case of \textit{choice G $\times$ E}, an interaction between $G_{i}$ and $E_{i}$ can arise if the effort-enhancing features of better schools are more successful at inducing effort among individuals with high levels of $G_{i}$ who already have a high propensity to supply effort. Alternatively, it can arise if the increased effort due to the reduced cost for high $G_i$ pupils is more productive in effort-enhancing schools. This is distinct from the other $G \times E$ channels because it takes place entirely at the level of $x_i$.  The productivity of effort could be identical across different levels of $G_{i}$ and $E_{i}$, but they could still interact in determining how much effort an individual exerts.       

The final term in \autoref{eq:GEDecomp} captures a gene--environment interaction that arises from nonlinearities in the production function, $F_{xx} \neq 0$ (e.g., diminishing returns to effort). For example, higher levels of $G_i$ may reduce the marginal product of school quality in producing $Y_{i}$. Students with high levels of $G_i$ may already be putting in long hours of study (regardless of the quality $E_i$ of their school), and this may reduce the marginal effect of improving $E_i$ if one of the mechanisms through which better schools operate is to encourage more effort.

\paragraph{Gene--environment interplay in welfare:} A common motivation for the study of \textit{G $\times$ E} interplay is to understand whether environmental factors dampen or amplify disparities in economic outcomes resulting from genetic endowments. A formal economic model of \textit{G $\times$ E} interplay helps clarify the conditions under which we can expect a difference between gene--environment interplay in observable outcomes $Y_{i}$ and in the welfare of decision-makers. Let $V=F(x_{i}^{*},G_{i},E_{i},e_{i})-C(x_{i}^{*},G_{i},E_{i},e_{i})$ represent an individual's value function, i.e., the maximum (optimized) value associated with \autoref{eq:valuefunction} above.  We can differentiate $V$ with respect to $E_i$ and $G_i$ to obtain an expression analogous to \autoref{eq:GEDecomp} but at the level of individual welfare:
\begin{eqnarray}
\frac{\partial^{2} V_{i} }{\partial G_{i} \partial E_{i}} &=&\frac{\partial^{2} Y^{*}_{i} }{\partial G_{i} \partial E_{i}} - \frac{\partial^{2} C^{*}_{i} }{\partial G_{i} \partial E_{i}}.  \label{eq:GEWelfareDecomp}
\end{eqnarray}

Specifically, we can derive the following expression for the difference between these two:

\begin{eqnarray}
\frac{\partial^{2} V_{i} }{\partial G_{i} \partial E_{i}}-\frac{\partial^{2} Y^{*}_{i} }{\partial G_{i} \partial E_{i}} &=&- \frac{\partial^{2} C^{*}_{i} }{\partial G_{i} \partial E_{i}} \notag \\
                                                          &=& -\left(C_{xx}x_{i,G}x_{i,E}+C_{xG}x_{E}+C_{xE}x_{G}+C_{x}x_{EG}+C_{EG}\right) \label{eq:GEWelfareDelta}
\end{eqnarray}

\autoref{eq:GEWelfareDelta} shows that the magnitude of \textit{G $\times$ E} interactions in the production of an outcome (like educational attainment) can either understate or overstate the extent of $G \times E$ interaction in the \textit{welfare} of individual decision-makers. For example, higher quality schools might induce individuals with lower levels of $G$ to reduce their own effort (as effort may be more costly for them than for individuals with higher levels of $G$). The lower level of $x$ might perfectly cancel out the added productivity from better tutors, leading to no effect on $Y$. However, in this case, students with lower $G$ do benefit from the higher quality environment. Without compromising their educational attainment, due to higher-quality tutors, students with lower $G$ can now increase their utility through reducing effort. This is captured in the above expression by a larger negative value for the term $-C_{x}x_{EG}$. That is, the increase in $E$ might cause a reduction in effort that is larger for individuals with a low level of $G$ ($x_{EG}>0$, while $x_{E}<0$). Then, if $C_{x}$ is a sufficiently large positive term, overall, we would have $\frac{\partial^{2} V_{i} }{\partial G_{i} \partial E_{i}}<0$, meaning that the policy might increase utility to a greater extent for individuals with lower genetic endowments $G$ even if it results in no differential change in $Y_{i}$. Thus, even relatively simple choice problems can substantially complicate the data generating process linking $G_{i}$, $E_{i}$, and the outcome of interest $Y_{i}$. Formally modeling these choice problems can guide the empirical analyst in understanding which variables to include in the analysis (inputs such as school quality, choices such as effort, outcomes such as grades, wages, or well-being) and the implications of $G \times E$ interplay in each of these variables.

\paragraph{Toward an empirical specification:} To link our theoretical model of $G \times E$ interplay to an empirical specification, we now consider a simple case in which the production function $F_i(\cdot)$ for $Y_{i}$ is modeled as a linear function of environmental factors $E_i$, genetic factors $G_i$, the investment choice $x_i$, their interactions, and an additive error $e^{f}_{i}$:
\begin{equation}
    F(x_{i},G_{i},E_{i},e_{i})=f_{x}x_{i}+f_{e}E_{i}+f_{g}G_{i}+f_{xe}x_{i}E_{i}+f_{xg}x_{i}G_{i}+f_{ge}G_{i}E_{i}+e^{f}_{i}.
    \label{eq:linear}
\end{equation}
In this linear framework, individuals choose an optimal level of investment $x_{i}$ to maximize utility, which we assume is simply the level of $Y_{i}$ net of quadratic costs of investment:
\begin{equation}
\max_{x_{i}}\:Y_{i}-\frac{c}{2}x_{i}^{2}.
\end{equation}
These choices guarantee that the second-order condition is met. An optimal choice for $x_{i}^{*}$ solves the following first-order condition:
\begin{equation}
    x_{i}^{*}=\frac{1}{c}\frac{\partial\:Y_{i}}{\partial\:x_{i}}(x_{i}^{*})=\frac{1}{c}\left[f_{x}+f_{xe}E_{i}+f_{xg}G_{i}\right]. \label{eq:xstar}
\end{equation}
Thus, optimal effort $x_i^*$ is higher when higher effort translates into improved outcomes (e.g., educational attainment; $f_x>0$) and when both environmental factors $E_i$ (e.g., school quality) and genetic factors $G_i$ (genetic propensity toward schooling) reinforce effort $x_i^*$, i.e., when $f_{xe}>0$ and $f_{xg}>0$ (and nonnegligible) and when the cost of effort $c$ is small. Effort is not affected by the terms $f_e$, $f_g$ and $f_{ge}$, as these reflect the ``technology'' of the outcome and operate independently of endogenous effort $x_i$. 

We now make the assumption that the (inverse) marginal cost of $x_{i}$ is a function of $E_{i}$ and $G_{i}$:
\begin{equation}
    \frac{1}{c}=k_{o}+k_{e}E_{i}+k_{g}G_{i}+e^{k}_{i}.
\end{equation}
Here, $e^{k}_{i}$ is an unobserved, idiosyncratic factor affecting the marginal cost of $x_{i}$. Substituting this expression for $\frac{1}{c}$ into \autoref{eq:xstar}, and our expression for $x^{*}_{i}$ into the production function \autoref{eq:linear}, yields an expression for the endogenously determined outcome $Y_{i}$ as a function of model primitives:
\begin{eqnarray}
Y_{i}&=&z_{0}f_{x}+\left[f_{g}+f_{x}z_{g}+z_{0}f_{xg}\right]G_{i}+\left[f_{e}+f_{x}z_{e}+z_{0}f_{xe}\right]E_{i}+ \nonumber  \\ 
     && \:\:\:\:\: \left[f_{ge}+f_{xe}z_{g}+f_{xg}z_{e}\right]G_{i}E_{i}+\left[f_{x}z_{g2}+f_{xg}z_{g}\right]G_{i}^{2}+\left[f_{x}z_{e2}+f_{xe}z_{e}\right]E_{i}^{2}+ \nonumber \\
     && \:\:\:\:\: f_{xg}z_{g2}G_{i}^{3}+f_{xe}z_{e2}E_{i}^{3}+\left[f_{xg}z_{ge}+f_{xe}z_{g2}\right]\left(G_{i}^{2}\times{}E_{i}\right)+\left[f_{xe}z_{ge}+f_{xg}z_{e2}\right]\left(G_{i}\times{}E_{i}^{2}\right)+\nonumber \\
     && \:\:\:\:\: \left(e^{f}_{i}+e^k_{i}\left[f_{x}+f_{xe}E_{i}+f_{xg}G_{i}\right]\left[f_{x}+f_{xe}E_{i}+f_{xg}G_{i}\right]\right), 
     \label{eq:endogenousY}
\end{eqnarray}

This can be simplified to 
\begin{eqnarray}
Y_{i}&=&\alpha+\beta_{G} G_{i}+ \beta_{E} E_{i}+ \beta_{G\times E} \left(G_{i} \times E_{i}\right) +\beta_{G2}{}G_{i}^{2}+\beta_{E2}{}E_{i}^{2}+ \nonumber \\
 && \:\:\:\:\:\beta_{G3}{}G_{i}^{3}+\beta_{E3}{}E_{i}^{3}+\beta_{G2E}\left(G_{i}^{2}\times{}E_{i}\right)+\beta_{GE2}\left(G_{i}\times{}E_{i}^{2}\right)+\epsilon_{i},
\label{eq:gxe}
\end{eqnarray}
where the coefficient for each variable on the right-hand side is composed of a mix of structural parameters that represent direct or indirect effects and interactions mediated by optimal choices (behavioral responses).

This linear model highlights the necessity of considering endogenous behavioral responses when empirically modeling $G \times E$ interplay or interpreting estimated $G \times E$ effects. In this stylized setting, if all individuals make the same choice $x$, if $x_{i}$ is independent of $E$ and $G$, or if the costs of investments are high (i.e., the $z$ parameters are small), then one returns to the exogenous world in which behavioral adjustments $x_i$ in response to genetic endowments and environments are negligible, i.e., where effectively $f_x=0$, $f_{xe}=0$ and $f_{xg}=0$. In this case, the production function in \autoref{eq:linear} would suggest a relatively simple linear data-generating process with $G_{i}$, $E_{i}$, and $G_{i}\times{}E_{i}$ terms. Indeed, this is the standard specification in the literature \citep[e.g.,][]{Keller2014, Schmitz2017}.

However, when $x_{i}$ is endogenously determined by optimizing behavior, the data-generating process becomes substantially more complicated. \autoref{eq:gxe} features quadratic and cubic terms in both $G_{i}$ and $E_{i}$, along with the higher-order interaction terms $\left(G_{i}^{2}\times{}E_{i}\right)$ and $\left(G_{i}\times{}E_{i}^{2}\right)$. Hence, optimal behavior implies that the model describing the relationship of $Y_i$, $G_i$, and $E_i$ should include the nonlinear terms $E_i^{2}$ and $G_i^{2}$ even when the production function does not include them. Moreover, even in the absence of an interaction between genes and environment in the production function ($f_{ge}=0$), interactions may arise purely through behavioral responses: $f_{xe}z_g + f_{xg}z_e$. Modeling these dynamics is of crucial importance for the correct estimation and interpretation of $G \times E$ interplay.

Finally, the structure of the error term $\epsilon_{i}$ in \autoref{eq:gxe} is informative:
\begin{equation}
    \epsilon_{i}=e^{f}_{i}+e^k_{i}\left[f_{x}^{2}+2f_{x}f_{xe}E_{i}+2f_{x}f_{xg}G_{i}+2f_{xe}f_{xg}\left(G_{i}\times{}E_{i}\right)+f_{xe}^{2}E_{i}^{2}+f_{xg}G_{i}^{2}\right]
\end{equation}
Here we see that in the presence of heterogeneity in the marginal investment costs ($e^k_{i}$), we essentially have a random coefficients model, with $E_{i}$, $G_{i}$, and $\left(G_{i}\times{}E_{i}\right)$ all entering multiplicatively into the error term. This means that heterogeneity in $e^k_{i}$ necessarily induces heteroskedasticity in the error term $\epsilon_{i}$. 

\section{Measuring \textit{G}}
\label{sec:G}
\subsection{Genetics in a nutshell}
Human DNA is composed of sequences of approximately 3 billion pairs of nucleotide molecules. These nucleotides come in four varieties: adenine (A), guanine (G), cytosine (C) and thymine (T). The nucleotides together constitute the genome. The human genome is divided into 23 pairs of chromosomes (22 so-called autosomal chromosomes and 1 sex chromosome), where for each pair, one chromosome is inherited from the mother and one from the father. Each chromosome contains a single double-stranded piece of DNA. ``A'' on one strand is always paired with ``T'' on the other strand, and ``C'' is always paired with ``G''. These combinations are called base pairs, and stretches of these pairs coding for a protein are called genes. The human genome consists of around 25,000 genes that code for proteins with a specific function \citep{sequencing2004finishing}. In addition, there are regions in-between genes with important regulatory functions. 

Most nucleotides ($\sim$99.9\%) in human DNA are identical from person to person. The part of DNA where people are different from each other are called polymorphisms and the location of most polymorphisms are well known. For many applications, it is therefore not necessary to sequence each individual's full genome. The most common polymorphism is a Single Nucleotide Polymorphism (SNP), where there is variation at a single-nucleotide locus. These variants of nucleotides are called alleles. In the human genome, there are approximately 85 million SNPs with a minor allele (i.e., the less common allele) frequency of $>1\%$ \citep{1000Genomes2015}. Current genotyping arrays measure several million of these SNPs, and many more that are not measured (typically $>$40 million) can be imputed with high accuracy because of the correlation structure in the genome \citep[so-called linkage disequilibrium,][further explained below]{reich2001linkage} and the availability of large reference panels \citep{quick2019sequencing}. 
SNPs have been the focus of most genetic discovery studies in the literature, and we follow this precedent. It is common to quantify SNPs by counting the number of minor alleles. Hence, at each locus, a SNP can take the values 0, 1 or 2. Individuals who inherited the same allele from each parent are called homozygous for that SNP (values 0 or 2), while individuals who inherited different alleles are called heterozygous (value 1). 

Virtually all outcomes that social science researchers are interested in are highly ``polygenic'' \citep{visscher2008heritability}. That is, there is no ``gene for'' a certain outcome, but individuals rather fall somewhere on a scale of genetic risk or predisposition that reflects the aggregation of numerous small contributions of millions of genetic loci. 

\subsection{Genome-wide association studies}
The polygenic nature of most traits was established through genome-wide association studies (GWASs). In a GWAS, one tests for associations between $J$ genetic variants (SNPs) and an outcome $y$ of interest without restricting the set of SNPs on theoretical grounds. Specifically, an ideal GWAS relates all SNPs ($x_{ij}$, coded as 0, 1, or 2, reflecting the number of minor alleles) to a specific outcome ($Y_i$) for individual $i$ in a regression framework of the form
\begin{eqnarray}
Y_i = \sum_{j=1}^{J} \beta_j x_{ij} + z_i^{'} \gamma + \epsilon_i,
\label{eq:gwas}
\end{eqnarray}
with SNP effects $\beta_j$, relevant controls $z_i$, and an error term $\epsilon_i$. In practice, however, this ideal model cannot be identified since existing datasets cover fewer individuals than SNPs \citep{Benjamin2011}.\footnote{At the time of writing, the biggest sample size of a GWAS is 5.4 million \citep{Yengo2022}.} 
GWASs therefore consider all $J$ SNPs by running sequential regressions for each SNP $j$, one at a time. Thus, in its most basic form, a GWAS regresses the outcome of interest on a single SNP $j$ and repeats this procedure $J$ times for every SNP: 
\begin{eqnarray}
y_i = \beta_j^{GWAS} x_{ij} + z_i^{'} \gamma_j + \epsilon_{ij}.
\label{eq:gwas2}
\end{eqnarray}
This produces a list of $\beta_j^{GWAS}$ coefficients for all $J$ SNPs. The set of control variables $z_i$ is usually very sparse, typically including age and sex alongside the first (usually ten) principal components (PCs) of the genetic data to account for population stratification \citep{Price2006}.\footnote{\label{fn:popstrat} Population stratification is a form of confounding where the genetic makeup of ancestors influences one's genetic makeup as well as an outcome through nongenetic pathways. More specifically, if a population is stratified into subpopulations that do not mate randomly and an outcome happens to be more common in one subpopulation for nongenetic reasons, then the outcome will appear to be correlated with any SNPs that also happen to be more common in that subpopulation. A commonly used hypothetical example is the ``chopstick gene'' \citep{hamer2000beware}, where people of Asian descent have different allele frequencies and tend to eat with chopsticks for cultural reasons. A GWAS investigating the genetic basis of chopstick use without controlling for ancestral differences in allele frequencies would then pick up a chopstick gene.} Importantly, due to the sparsity of control variables and the correlation between closely spaced SNPs, $\beta_j^{GWAS}$ is not necessarily equal to $\beta_j$.

The set of $\beta_j^{GWAS}$ coefficients is a simple linear projection of the outcome of interest on the space spanned by the measured SNPs. Imposing linearity neglects any form of interaction, whether gene--gene or gene--environment, thereby implicitly assuming that such interactions are negligible, or of second-order importance. From the perspective of the model introduced in \autoref{sec:econModel}, these $\beta_j^{GWAS}$ coefficients estimate the unconditional association between each SNP and the outcome $Y$, holding all other genetic and environmental factors constant at the sample average: $\beta_j^{GWAS} = {\partial Y(G,E,e)}/{\partial SNP_j}\Bigr|_{G=\bar{G},E=\bar{E},e=\bar{e}}$. Therefore, SNPs that have a constant effect across different environments are more likely to be identified in a GWAS than SNPs that have diverging (opposite-sign) effects in different settings. In other words, significant coefficients in GWAS are more likely to be detected for SNPs that do not have a sizable gene--environment interaction \citep[e.g.,][]{Mills2020book}.

Running so many regressions requires a correction for multiple hypothesis testing. Considering that there are $\sim$1 million \textit{independent} SNPs in the human genome (adjacent SNPs are often in linkage disequilibrium, i.e., inherited together), the commonly used criterion for genome-wide statistical significance is $p <5 \times 10^{-8}$ (i.e., 0.05 divided by 1,000,000). The stringent significance level, in combination with the tiny effect sizes of individual SNPs on outcomes \citep{Rietveld2013,Chabris2015}, necessitates the use of extremely large samples to ensure adequate power. Legal and privacy reasons usually prohibit the joint analysis of genetic datasets. For this reason, researchers typically pursue a meta-analysis strategy to obtain a sufficiently large analysis sample \citep{Visscher2017}. Consortia such as the Social Science Genetic Association Consortium (SSGAC), Genetic Investigation of ANthropometric Traits (GIANT) and GWAS \& Sequencing Consortium of Alcohol and Nicotine Use (GSCAN) have been key to this, coordinating the analyses (harmonizing the outcomes, quality-controlling consortium datasets) and bringing together the results of large numbers of smaller datasets. In such meta-analyses, only GWAS summary results (the effect sizes for each SNP) are shared between consortium members, addressing the legal and privacy barriers to their joint use. 
The GWAS meta-analysis approach has made possible an unprecedented surge in genetic discoveries that replicate consistently \citep{Visscher2017}.\footnote{By contrast, so-called candidate-gene studies, a hypothesis-driven approach, have weak replication records \citep{Hewitt2012,Chabris2013}. } 

\begin{figure}
    \centering
    \includegraphics[width=14cm]{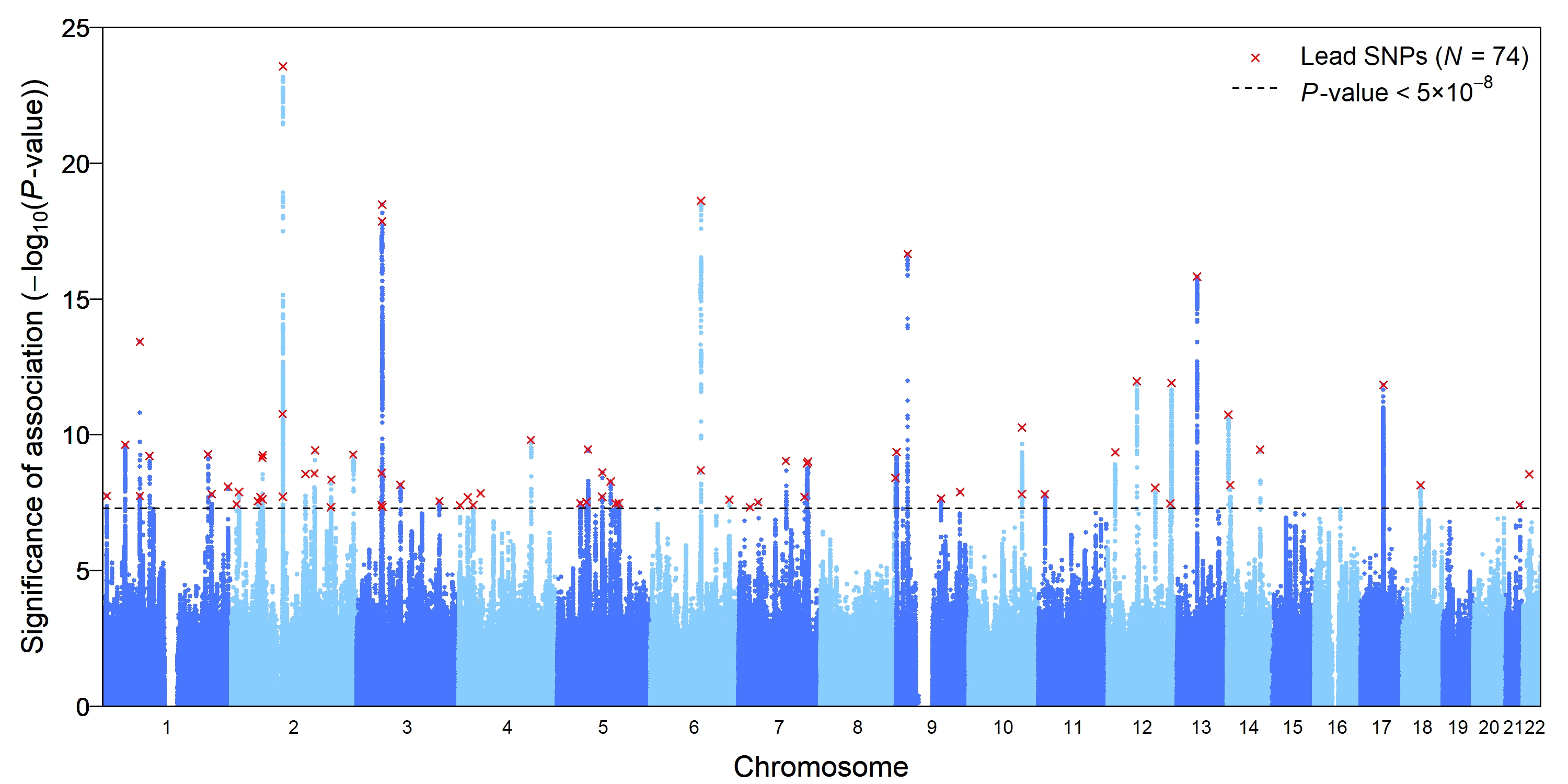}
    \caption{Manhattan plot visualizing the results of the second genome-wide association study on educational attainment by \cite{Okbay2016}.}
    \label{fig:manhattan}
\end{figure}

GWAS results are typically presented using so-called Manhattan plots. As an example, \autoref{fig:manhattan} provides the Manhattan plot visualizing the results of the second GWAS of educational attainment \citep{Okbay2016}. The $x$ axis of the Manhattan plot represents the position of the SNP in the genome (the numbers 1-22 reflect the autosomal chromosomes) and the $y$ axis the strength of the evidence for an association with the outcome variable (as reflected in the $p$ value). The $p$ value is transformed (by taking the negative of the 10 log of the $p$ value) so that higher values represent stronger associations. Specifically, when a dot (representing a single SNP) is above the dashed line in the Manhattan plot (note that –log$_{10}(5\times10^{-8}) = 7.3$), the SNP is genome-wide significant. The Manhattan plot also visualizes the effect of linkage disequilibrium. Because SNPs physically close to one another are more likely to be inherited together (i.e., are in linkage disequilibrium), the regression results are very similar for adjacent SNPs. As a result, $p$ values of adjacent SNPs are highly correlated. This is visible in the towers of dots around genome-wide significant SNPs. The result looks like the skyscrapers of Manhattan towering above lower-level buildings.

Because of linkage disequilibrium, each genome-wide significant SNP is correlated with adjacent SNPs. Each \textit{block} of correlated SNPs is called a genome-wide significant locus. One typically picks a lead SNP, the SNP in a genome-wide significant locus with the smallest $p$ value. By construction, the set of lead SNPs are therefore approximately uncorrelated with each other. The very first GWAS of educational attainment used a sample of $\sim$125,000 people \citep{Rietveld2013} and identified 3 genome-wide significant loci. The second GWAS of educational attainment used a sample of $\sim$400,000 people \citep{Okbay2016} and identified 74 genome-wide significant loci. The third, \cite{lee2018gene}, used a sample of $\sim$1.1 million individuals to uncover 1,271 lead SNPs, and the fourth GWAS of educational attainment used $\sim$3 million individuals and identified 3,952 lead-SNPs associated with educational attainment \citep{Okbay2022}. This rapid growth in the power of genetic discovery exemplifies the genetics revolution that we are in the midst off. 

\subsection{Polygenic indices}
\label{sec:polygenicindices}
The tiny explanatory power of individual SNPs has led researchers to develop methods which combine individual SNPs into so-called polygenic indices (PGIs), which have substantially greater explanatory power. A PGI is a weighted sum of individual SNPs and reflects the best linear genetic predictor of an outcome \citep[e.g.,][]{Mills2020book,becker2021resource}. It is constructed with the aim of predicting the genetic propensity toward a certain trait for individuals in a hold-out sample. For reasons of statistical independence, the hold-out sample cannot have been part of the original GWAS meta-analysis. 

In its most basic form, a PGI is constructed as follows: 
\begin{eqnarray}
PGI_i & = & \sum_{j=1}^{J} \beta_j^{GWAS} x_{ij},
\label{eq:PGI}
\end{eqnarray}
where $x_{ij}$ is again the number of copies of the minor allele for individual $i$ and SNP $j$ and {\small $\beta_j^{GWAS}$} are the $\beta$ coefficients for SNP $j$ (see \autoref{eq:gwas2}) from the corresponding GWAS  \citep{dudbridge2013power}. By multiplying SNP $j$ (taking values $x_{ij} = \{0,1,2\}$) with its {\small $\beta_j^{GWAS}$} weight, SNPs with large effect sizes are weighted higher than those with small effect sizes. The simplest PGIs follow \autoref{eq:PGI} where, given the linkage disequilibrium (LD) between SNPs, only one out of each genome-wide significant locus is maintained in the computation of the PGI. More sophisticated measures exist that account directly for LD \citep[see, for example,][]{so2017improving,Vilhjalmsson2015}, with typically better predictive power. However, all approaches have in common that they aggregate the genetic contributions of millions of small SNP-effects across the genome and are similar in spirit to the basic (and still commonly used) approach of the linear weighted sum in \autoref{eq:PGI}.

PGIs that include all available SNPs (i.e., genome-wide significant as well as nonsignificant SNPs) typically explain most variation in the outcome \citep{Ware2017}. The predictive accuracy of a PGI is also an increasing function of the sample size of the GWAS \citep{dudbridge2013power}. As GWAS samples grow, the estimates of the $\beta_j^{GWAS}$ coefficients improve, and measurement error in the PGI is reduced. For example, whereas the PGI based on the first successful GWAS on educational attainment ($N \sim 125,000$) explained 3-4\% of the variance in educational attainment out-of-sample \citep{Rietveld2014}, the PGI based on the results of a second GWAS \citep[][$N \sim$ 400,000]{Okbay2016} explained 6-8\%, the PGI based on a third GWAS \citep[][$N \sim $ 1,100,000]{lee2018gene} explained 11-13\%, and the PGI based on the fourth GWAS \citep[][$N \sim $ 3,300,000]{Okbay2022} explained 13-16\% of the variation in educational attainment. The maximum explained variance of a PGI is determined by so-called SNP-based heritability. Using methods like Genome-based Restricted Maximum Likelihood (GREML) estimation \citep{Yang2011}, several studies have shown that this number is around 25\% for educational attainment \citep{Rietveld2013}. In other words, today's PGI for educational attainment already explain a bit more than half the variation in educational attainment that is thought to be achievable. 

\subsection{The endogenous nature of polygenic indices} 
\label{subsec:interpretG}
While PGIs constitute the best linear genetic predictors of an outcome, it is important to emphasize that this holds \textit{within the environmental and demographic context of the discovery sample} \citep{Mills2020book, domingue2020interactions}. Thus, the association between a PGI and an outcome cannot be interpreted as an immutable biological relationship \citep[e.g.,][]{Mostafavi2017,kweon2020genetic}: the effects depend on the context, i.e., on the environment. As PGIs are constructed using effects estimated in a GWAS, environmental factors may influence the PGI through at least three channels: as moderators, confounders, and mediators. First, as a \underline{moderator}, the environment could change the strength of the relationship between a PGI and the outcome. This is precisely the topic of this paper, namely, the interplay between genes $G$ and environment $E$; environmental moderation would be reflected by an interaction term $G \times E$. The role of environmental factors as confounders and mediators requires some further explanation. 

\begin{figure}[ht]
    \centering
    \includegraphics[width=14cm]{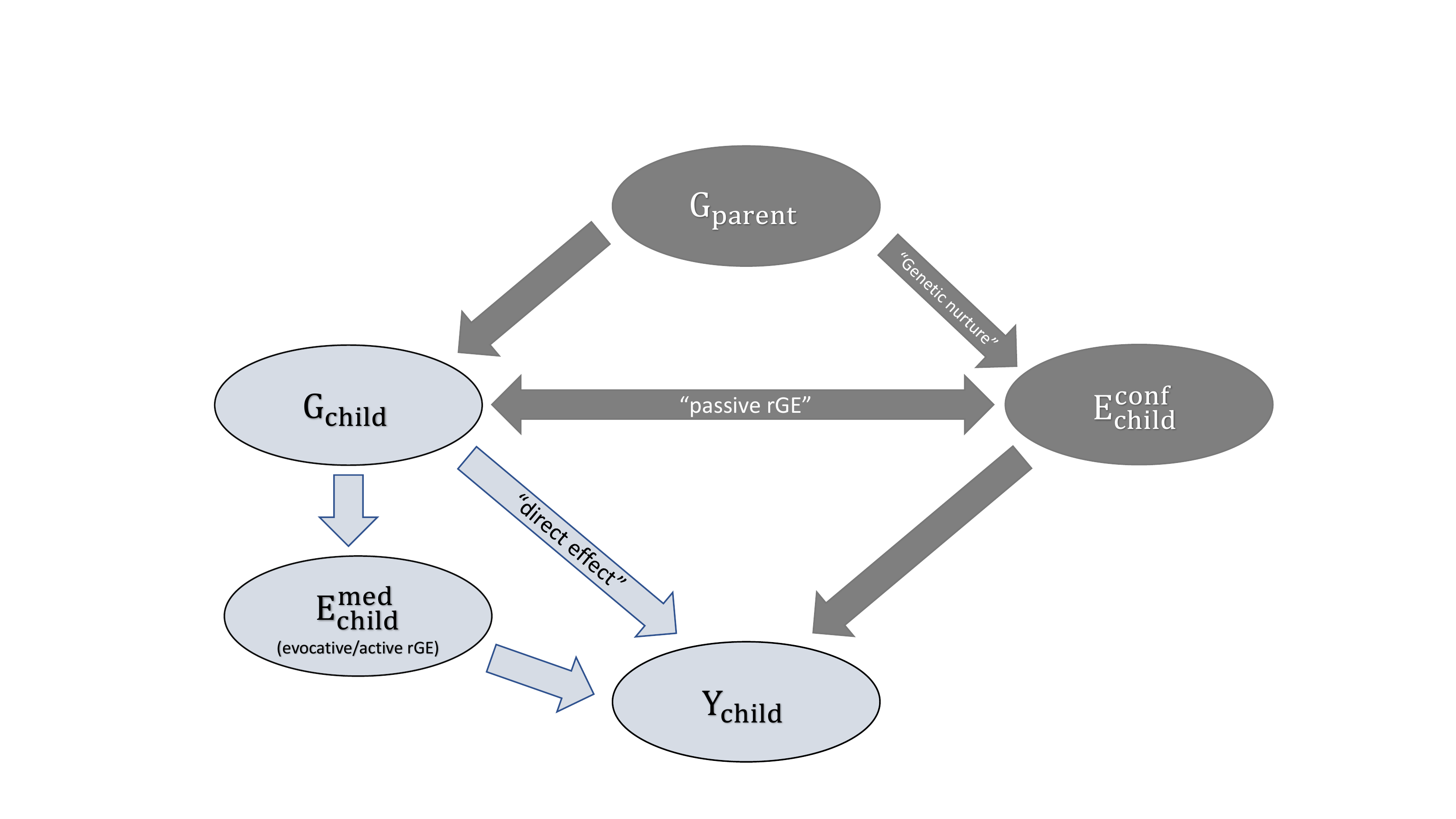}
    \caption{Diagram visualizing the relationship between parental genes ($G_{parent}$), the child's genes ($G_{child}$), environmental factors ($E_{child}^{med}$ and $E_{child}^{conf}$), and the outcome ($Y_{child}$). The darker area represents the ``causal'' region of the diagram (explained further in the text).}
    \label{fig:diagram}
\end{figure}

To fix ideas, we again assume that our outcome variable is educational attainment.  \autoref{fig:diagram} shows a schematic of the relationships between the parental genotype ($G_{parent}$), the child's environment shaped by parents ($E_{child}^{conf}$), the child's genotype ($G_{child}$) and the child's outcome ($Y_{child}$). The top part of the diagram reflects the basic notion that a child inherits her genotype from her parents (arrow from $G_{parent}$ to $G_{child}$). The child's genotype in turn may have a direct effect on the outcome (arrow from $G_{child}$ to $Y_{child}$, labeled ``direct effect'').\footnote{As mentioned before, genetic effects depend on the environmental context: they are not immutable or deterministic. For example, even though alcohol metabolism and dependence are found to be partially due to genetic factors, in an environment where drinking alcohol is illegal, the direct effects would be zero.} At the same time, the genotypes of parents translate into certain environments, i.e., parents with genotypes conducive to education may provide an environment more beneficial to their child's learning (so-called genetic nurture; arrow from $G_{parent}$ to $E_{child}^{conf}$). This environment in turn may raise the child's educational achievement (arrow from $E_{child}^{conf}$ to $Y_{child}$). The child's environment $E_{child}^{conf}$ here acts as a \underline{confounder} (hence the superscript \textit{conf}). This is because $E_{child}^{conf}$ not only influences the outcome $Y_{child}$ but also is correlated with the child's genotype $G_{child}$. The existence of genetic nurture has been demonstrated by significant associations between \textit{nontransmitted} parental genotypes and children's outcomes \citep{Bates2018,Kong2018,Wertz2018}. Since the child did \underline{not} inherit these genetic variants, this correlation must operate through environmental channels.
 Genetic nurture is an example of so-called passive gene--environment correlation ($rGE$), which occurs when individuals' genotypes are related to their environment but that environment is \underline{not} a consequence of the child's genotype $G_{child}$.\footnote{Other sources of passive $rGE$ exist, including population stratification (see footnote \ref{fn:popstrat}). 
Passive $rGE$ also occurs when siblings' genotypes partially shape the environment that individuals are exposed to \citep[see, e.g.,][]{cawley2019testing}. Since siblings (like parent--child pairs) share, on average, 50\% of their DNA, this introduces a correlation between sibling genotypes and sibling environments. Last, passive $rGE$ can also arise from assortative mating. If phenotypic selection (for example, based on education) induces greater genetic similarity between partners than in the general population, this will lead to biased estimates of the causal effect of a genotype on a phenotype in subsequent generations \citep{morris2020population}. In short, passive $rGE$ arises from several sources besides parental genotype \citep{Mills2020book}, yet conditional on parental genotype, the child's genotype is as good as random, fully eliminating any confounding due to passive $rGE$.} 
Passive $rGE$ is reflected by the horizontal double arrow between $E_{child}^{conf}$ and $G_{child}$.

In general, gene--environment correlation ($rGE$) describes the phenomenon of certain environments being more prevalent among carriers of certain genotypes \citep{Plomin1977,Fletcher2013}. Two other types of $rGE$ are generally considered: evocative and active $rGE$ \citep{Plomin1977}. 
First, \textit{evocative} gene--environment correlation occurs when someone's genetic predisposition $G_{child}$ invokes a certain environmental response; $E_{child}^{med}$ (lighter area of the diagram). For example, a child with a high genetic predisposition for calmness may be treated more favorably by her parents and teachers, creating an environment that may be more conducive to learning.

Second, \textit{active} gene--environment correlation occurs when individuals with certain genotypes $G_{child}$ purposefully self-select into certain environments $E_{child}^{med}$. For example, someone with a high genetic predisposition for education may find it easier to apply for and be accepted into a selective, high-quality university. Hence, both active and evocative $rGE$ imply that the environment $E_{child}^{med}$ is a \textit{consequence} of the child's genotype $G_{child}$. These environments, in turn, influence the child's educational attainment $Y_{child}$. Hence, through active and evocative $rGE$, the environment may act as a \underline{mediator}, a variable that is influenced by the child's genotype $G_{child}$ and that in turn influences the outcome $Y_{child}$ (hence the superscript \textit{med} in $E_{child}^{med}$).

Genotypes have the useful property of being fixed at conception, and therefore, the outcome cannot affect the genotype (i.e., there is no reverse causality). We adopt the view that the causal effect of genotype $G$ can be thought of as a \textit{variant substitution} effect \citep{lee2013causal,morris2020population}. That is, the causal effect of a genetic variant is the counterfactual change in an individual's outcome that would occur had that genetic variant been different at conception, with all else held constant.\footnote{Other definitions, for example that of \citet{young2020mendelian}, also include indirect genetic effects stemming from passive $rGE$ as part of the causal effect of $G$. From a dynastic point of view, these definitions are similar---hypothetically, a change in one's genotype at conception will have implications for individuals and their offspring (i.e., lead to passive $rGE$ for the next generation). However, here we take an individual's genotype and her life cycle as the relevant unit of analysis and therefore treat passive $rGE$ arising from relatives' genotypes as a source of bias rather than a causal effect.} 
In the diagram, the mechanisms through which the effect of the child's genotype $G_{child}$ operate can be through direct pathways (e.g., gene expression) but can also be environmentally driven (e.g., through active or evocative $rGE$). These causal genetic pathways from $G_{child}$ to the child's outcome $Y_{child}$ fall within the (lighter) causal part of the diagram. The existence of (passive) $rGE$ implies that the offspring's genotype $G_{child}$ and outcome $Y_{child}$ are simultaneously influenced by the parental genotype $G_{parent}$. This leads to endogeneity of $G_{child}$, if not properly controlling for for parental genotype.

As we discuss in detail in \hyperref[subsec:ideal]{Section~\ref*{subsec:ideal}}, controlling for parental genotype can fully address confounding due to passive $rGE$, allowing researchers to make causal inferences. This is because, conditional on the genotype of parents $G_{parent}$, the genotype of the child $G_{child}$ is as good as random (``Mendel's Law''), breaking the link between $G_{child}$ and child environment $E_{child}^{conf}$ (see \autoref{fig:diagram}).

In the next section, we discuss approaches to addressing the endogeneity of the PGI and the environment in analyses of $G \times E$ interplay and the more general question of how to estimate any \textit{moderating} effect of an environment on a genotype.

\section{An empirical specification of \texorpdfstring{$G \times E$}{} interplay}
\label{sec:specification}
The core idea behind gene--environment interplay ($G \times E$) is that nature and nurture are not additive and separable but intrinsically joined and nonlinear. Interaction effects---a concept that appears in several disciplines---may also be referred to as synergies, complementarities, supermodularity, or heterogeneity of treatment effects \citep{Mullahy1999,Mullahy2008}. Following the economic model in \hyperref[sec:econModel]{Section~\ref*{sec:econModel}}, let us again consider a data-generating process in which an outcome $Y_{i}=F(x^{*},G_{i},E_{i},e_{i})$ is a function of genetic endowments $G_{i}$, the environment $E_{i}$, random factors $e_{i}$, and optimal choices $x^{*}=x^{*}(G_{i},E_{i},e_{i})$.  Let us further assume that this data-generating process is additively separable in the random components $e_i$ so that it can be written as  

\begin{equation*}
    Y_i = \widetilde{F}(x_i,G_i,E_i) + \epsilon_i.
\end{equation*}

Note that here $\widetilde{F}(x_i,G_i,E_i)=E\left[F(x^{*},G_{i},E_{i},e_{i})\right]$ and $\epsilon_i$ is a mean-zero error term that is a function of $e_{i}$, $E_{i}$, and $G_{i}$ (as in the example given in \hyperref[sec:econModel]{Section~\ref*{sec:econModel}}). To test for the existence of $G \times E$, one needs to test for nonlinearities in the function $\widetilde{F}()$, specifically that $\partial^2 \widetilde{F}/\partial G \partial E \neq 0$.
We focus on the identification of $G \times E$ in the context of regression models. We start with the linear regression derived in \autoref{eq:gxe} in \hyperref[sec:econModel]{Section~\ref*{sec:econModel}}:
\begin{align}
Y_{i}=\alpha & +\beta_{G} G_{i} + \beta_{E} E_{i} + \beta_{G \times E} \left(G_{i} \times E_{i}\right) + \theta{}E_{i}^{2} + \rho{}G_{i}^{2} \nonumber \\
             & +\mu_{x} X_{i} + \mu_{g} \left(G_{i} \times X_{i}\right) + \mu_{e} \left(E_{i} \times X_{i}\right) +\varepsilon_{i}.
 \label{eq:function}
 \end{align}

Compared with \autoref{eq:gxe}, this empirical specification also includes control variables $X_{i}$ and full interactions between the control variables and the genetic and environmental measures. The interactions are included to ensure that the coefficient of interest $\beta_{G \times E}$ does not capture spurious correlations between $X_i$ and either $G_i$ or $E_i$ \citep[see][for details]{Keller2014}. \autoref{eq:function} does not include higher-order interactions involving $E^{2}$ and $G^{2}$ that are present in \autoref{eq:gxe}. As is customary in the literature, we exclude them here for simplicity but note that even simple, highly stylized economic models of $G \times E$ interplay as in section \ref{sec:econModel} suggest that empirical specifications should include even higher order terms than are typically estimated.

In \autoref{tab:Interpretation}, we present nine possible scenarios for estimating gene-by-environment interplay based on (exogeneity) assumptions for $G$ and $E$. In the first column, we distinguish between three possible scenarios of genotype $G$: 
(1) exogenous $G$ (i.e., family data are available, allowing one to control for parental genotype or include family fixed effects) \underline{and} a PGI obtained from a parent--child GWAS, (2) exogenous $G$ and a PGI based on a regular GWAS, and (3) endogenous $G$ (i.e., no family data are available) and a PGI based on a regular GWAS.\footnote{
There is a fourth category where the analyst has access to a PGI based on the results of a parent--child GWAS but applies this in a sample without family data. This case is currently extremely rare, although it might become more common after the publication of \cite{Howe2021}. However, since the analysis in this scenario uses a PGI without controls for parental genotype, the issues highlighted in scenario (2) also apply here. We therefore do not separately discuss this case.}

In the columns of \autoref{tab:Interpretation}, we distinguish between three categories for the environmental measure: exogenous and endogenous $E$, and an additional category of ``predetermined'' $E$. Predetermined measures of environment $E$ are defined as not caused by genotype $G$ but possibly correlated with \textit{other environmental characteristics} (which we refer to as $E^*$) or with parental genotype. Examples of predetermined environments $E$ could be family income or air pollution levels at the time of birth. Such environmental exposures are clearly not caused by one's genes but are likely to be correlated with other environmental exposures and possibly influenced by parental genotype.\footnote{We do not distinguish between predetermined and endogenous $G$ since $G$ is fixed at conception and therefore cannot be caused by subsequent measures of $E$ or $Y$. In other words, there is no reverse causality running from the outcome $Y$ or the environment $E$ to the genotype $G$.}

In the following subsections, we discuss each of the nine scenarios represented in \autoref{tab:Interpretation}. For reasons of space and exposition, we mostly focus on the interpretation and biases in the main effects of $G$ and $E$ and do not separately discuss the interaction term $G \times E$. However, in specific cases the interpretation of the interaction term does not follow naturally from the main effects and we will discuss those cases separately.

\begin{landscape}
\begin{table}[H] \caption{Estimation scenarios for $G \times E$ effects in gene--environment interaction models.} \label{tab:Interpretation}
\centering
{\scriptsize
\begin{tabular}{llll} \\
\hline \hline \\
& \multicolumn{1}{c}{Exogenous $E$} & \multicolumn{1}{c}{Predetermined $E$} & \multicolumn{1}{c}{Endogenous $E$} \\ 
\\
\hline
& & & \\

Exogenous $G$ (family data) \&       & $\checkmark G$ unbiased (causal) & $\checkmark G$ unbiased (causal) & $\checkmark G$ unbiased (causal) \\
\indent\hspace{0.3cm} PGI on basis of parent--child GWAS       & $\checkmark E$ unbiased (causal)        & $\uparrow \downarrow E$ may reflect (predetermined) $E^*$ through & $\uparrow \downarrow E$ may reflect $E^*$ through correlated environments\\
 &   & \indent\hspace{0.3cm} correlated environments  & \indent\hspace{0.3cm} or $G$ through active/evocative $rGE$ \\
&&&\\
\hline
&&&\\

Exogenous $G$ (family data) \&        & $\downarrow G$ downward biased (within-family measurement & $\downarrow G$ downward biased (within-family measurement & $\downarrow G$ downward biased (within-family measurement \\
\indent\hspace{0.3cm} PGI on basis of regular GWAS  & \indent\hspace{0.3cm} error \& overcontrol for genetic effect) & \indent\hspace{0.3cm} error \& overcontrol for genetic effect) & \indent\hspace{0.3cm} error \& overcontrol for genetic effect) \\
  & $\checkmark E$ unbiased (causal) & $\uparrow \downarrow E$ may reflect (predetermined) $E^*$ through & $\uparrow \downarrow E$ may reflect $E^*$ through correlated environments\\
&  & \indent\hspace{0.3cm} correlated environments  & \indent\hspace{0.3cm} or $G$ through active/evocative $rGE$ \\
&&&\\
\hline
& & & \\

Endogenous $G$ (no family data) \& & $\uparrow G$ upward biased; may reflect $E^*$ or parental $G$  & $\uparrow G$ upward biased; may reflect (predetermined) $E^*$   & $\uparrow G$ upward biased; may reflect $E$, $E^*$ or parental $G$\\
\indent\hspace{0.3cm} PGI on basis of regular GWAS
  & $\checkmark E$ unbiased (causal) & \indent\hspace{0.3cm} or parental $G$ & $\uparrow \downarrow E$ may reflect $E^*$ or parental $G$, or $G$ through\\

  &  & $\uparrow \downarrow E$ may reflect (predetermined) $E^*$ or parental $G$  & \indent\hspace{0.3cm} active/evocative $rGE$ 
 \\
&&&\\
\hline \hline
\end{tabular}}
\caption*{\footnotesize {\textit{Notes:} $G$ stands for genotype, $E$ for environment, $E^*$ for environments \textit{other than} those of interest, and $rGE$ for gene--environment correlation. A predetermined environment $E$ is defined as an environment not causally influenced by one's genes $G$ yet possibly correlated with other environmental characteristics $E^*$ and potentially shaped by parental genes. GWAS stands for genome-wide association study. In addition to the sources of bias presented in the table, any classical measurement error will lead to attenuation bias of the relevant parameter \textit{and} of the interaction parameter.}}
\end{table}
\end{landscape}

\subsection{The ideal experiment: Exogenous \textit{G} and exogenous \textit{E}}
\label{subsec:idealintro}

Observed environments are almost always endogenous. 
There could be a myriad of potential unobserved factors that influence the individual's environment and her outcome. To address this endogeneity, a useful starting point is to exploit exogenous sources of variation. For example, quasi-experimental designs have been used to isolate variation in environmental exposure that is independent of genotype and other potential confounders \citep[see, e.g.,][]{Schmitz2017,Barcellos2018}. Individuals' genotypes are also endogenous, e.g., due to passive $rGE$. Family designs allow us to address this endogeneity either by controlling for parental genotype or by including family fixed effects. We discuss this in more detail below.

If both the genotype $G$ and the environment $E$ are exogenous (a rare situation, as we discuss below, but one that is likely to arise in the not-too-distant future), the coefficients on $G$, $E$, and $G \times E$ are all unbiased in the $G\times E$ regression model. In \autoref{tab:Interpretation}, we distinguish two scenarios in which both $G$ and $E$ are exogenous. In the top left corner of   \autoref{tab:Interpretation}, we have a scenario in which we exploit an exogenous $E$ in combination with an exogenous $G$ and a PGI based on the results of a parent--child GWAS. In the cell beneath it, we have a situation with an exogenous $E$ in combination with an exogenous $G$ and a PGI based on the results of a regular GWAS. We first discuss these two scenarios in more detail.

\subsubsection{Exogenous \textit{G} and a PGI based on a parent--child GWAS}
\label{subsec:ideal}

The source of variation in one's genotype is well-understood. As stated by Mendel's first law, one's genes are the result of the random segregation of one's parental genes during meiosis. Thus, conditional on parental genotype, the genotype of the child is random. Controlling for parental genotype, therefore, would fully address the confounding that results from passive $rGE$, where parental genotype acts as a third variable that causes both the offspring's genotype and the offspring's environment \citep{bates2020causal}. Any association between a genetic variant and the outcome of interest uncovered by a GWAS that conditions on parental genotypes would reflect a causal genetic effect (following our definition of variant substitution). PGIs constructed from the results of such a family-based GWAS would represent the aggregated causal genetic effect for that outcome. When such PGIs in turn are applied in a family-based dataset that enables conditioning once again on parental genotypes, the coefficient on the PGI represents a causal genetic effect. 

Consider the following relation between the outcome $Y_i$ of the child $i$ and her genotype $G_i$, conditional on the genotype of her mother $G_m$ and of her father $G_f$ \citep{kong2020family}:
\begin{equation}
Y_i = \alpha + \delta G_i + \alpha_m G_{m(i)} + \alpha_f G_{f(i)}.
\label{eq:Kong1}
\end{equation}
Here, $\alpha$ is a constant term, and $\delta$ captures the direct genetic effect of the child's genotype $G_i$. The parameters $\alpha_m$ and $\alpha_f$ can be written as $\alpha_m = \eta_m +w$ and  $\alpha_f = \eta_f +w$, where $\eta_m$ and $\eta_f$ denote genetic nurturing effects from the mother and father, respectively \citep[see][]{kong2020family}, and $w$ captures all confounding effects that have not been adjusted for, including assortative mating, sibling interactions, and contributions from older ancestors. Since the correlation between the child's genotype $G_i$ and that of her parents is about 0.5, we can rewrite the previous equation as
\begin{equation}
Y_i = \alpha + \left(\delta + [\eta_m + \eta_f]/2 + w \right) G_i.     
\label{eq:Kong2}
\end{equation}

There are two useful and distinct ways of thinking about genotype $G$ in \autoref{eq:Kong1} and \autoref{eq:Kong2}: in terms of a GWAS stage and an analysis stage. We use these intermittently and somewhat loosely in what follows. The GWAS stage reflects a series of $J$ regressions that estimate $J$ beta effect sizes for each of $J$ SNPs, similar to \autoref{eq:gwas2} but now also conditioning on parental genotype (estimating not just $J$ $\delta$-effects but also $J$ $\alpha_m$- and $\alpha_f$-effects). The analysis stage represents the use of PGIs $G$ for the child and for the parents $G_m$ and $G_f$ based on results from various types of GWASs, and applied to a data set for analysis, e.g., a study of $G \times E$ interplay. As mentioned before, the PGIs should be based on GWASs that did not include the analysis sample for reasons of statistical independence.

Family-based GWASs that use parent--child dyads (called trios) literally regress \autoref{eq:Kong1} for each SNP $j$ of individual $i$ and those of her mother and father. By controlling for parental genotype, the genotype of the child is effectively randomized. Through such randomization, the child's genotype becomes orthogonal to the environment influenced by parents ($E_{child}^{conf}$), and the link between the blue (causal) and red (confounding) parts in  \autoref{fig:diagram} is broken. PGIs can be constructed for the offspring (child) in datasets that contain parent--child dyads (trios) by using summary statistics from such parent--child dyad GWAS results. Such PGIs are unbiased and can be interpreted as the causal effect of a genotype. When combined with an exogenous source of variation in the environment, such analyses would constitute the ideal experiment: when both $G$ and $E$ are exogenous, the estimated effects of the PGI $G$, the environment $E$, and the interaction between the PGI and the environment $G\times E$ will all be unbiased and can thus be interpreted as causal.

Whereas controlling for parental genotypes deals with the endogeneity of the offspring's genotype, in practice, there are currently no datasets with a sufficiently large number of parent--offspring trios to allow for a sufficiently powerful GWAS.\footnote{Imputation of the parental genotype on the basis of sibling genotypes or data from the other parent is a partial solution to this problem \citep{kong2020family,young2020mendelian}. This strategy requires genetic data for at least two family members in the same sample and has successfully been used to increase the effective number of parent--child dyads.} 
A common alternative strategy to establish causal effects of $G$ is to use a sample of sibling pairs and to run a \textit{within-family} analysis by including family fixed effects \citep{Howe2021}. Instead of \autoref{eq:Kong1}, we now have
\begin{equation}
Y_{ij} = \alpha_j + \delta G_{ij} + \varepsilon_{ij} ,
\label{eq:siblings}
\end{equation}
where $Y_{ij}$ is the outcome for individual $i$ in family $j$, $G_{ij}$ is the genotype of individual $i$ in family $j$, and $\alpha_j$ represents a family fixed effect absorbing the parental genotype.\footnote{Since the only confounding variables in a GWAS are the father's and mother's genotypes, the GWAS is sometimes also run as a regression without family fixed effects but with the mean sibling's genotype as a control variable \citep[e.g.,][]{Howe2021}. The mean sibling's genotype is a sufficient statistic to control for the mean influence of parental genotype and therefore leads to point estimates equivalent to those from \autoref{eq:siblings}.} 
The analysis compares differences in sibling genotypes $G_{ij}$ to differences between siblings in the outcome $Y_{ij}$ within families. Such analyses exploit the fact that the genotype variation between siblings is randomly assigned given that siblings draw from the same shared genetic pool: their parents. In many ways, the parent--child dyad approach and the family fixed effects approach are similar. When used in both the GWAS phase and the analysis phase, either approach delivers causal genetic effects (top left corner of \autoref{tab:Interpretation}) in the absence of sibling effects.

A clear advantage of family fixed effects strategies is that parental genotypes do not have to be observed, but there are three limitations of the family fixed effects approach relative to the parent--child dyad alternative. First, because the family fixed effects strategy requires at least two siblings from the same family, it cannot be used to study single-child families. Second, when one sibling's genotype directly affects another sibling's outcome, this will bias the coefficient of $G$ in a family fixed effects model \citep{kong2020family}.\footnote{For example, consider a case with two siblings where there is a direct effect $\gamma$ of one's sibling's genotype on the other siblings' phenotype:
\begin{align} Y_{1j} &= \alpha_j + \delta G_{1j} + \gamma G_{2j} + \varepsilon_{1j} \nonumber\\
 Y_{2j} &= \alpha_j + \delta G_{2j} + \gamma G_{1j} + \varepsilon_{2j}. \nonumber
 \end{align}
When taking sibling differences to eliminate the family fixed effects, we obtain
\[ Y_{1j}-Y_{2j} = \left(\delta - \gamma\right) \left(G_{1j}-G_{2j} \right) + \left(\varepsilon_{1j}-\varepsilon_{2j}\right). \]
When $\gamma$ is positive (negative), sibling effects cause a downward (upward) bias in the estimate of the effect of one's own genotype $G$, as measured by $\delta$.} %
In contrast, bias due to sibling effects does not exist in parent--child dyad analyses because these control for parental genotype. While there may still be sibling effects on the child's outcome, these no longer cause bias in the identified effect of the child's own genotype since the sibling genotypes are randomly assigned conditional on parental genotype, and hence independent of each other. A final limitation, specific to the context of $G \times E$, is that members of a sibling pair have to be exposed to different exogenous shocks to the environment. This is because the model is identified from variation between siblings within families. This puts strong restrictions on the nature of any natural experiment in a sibling approach to studying $G\times E$ interplay. In contrast, parent--child dyad analyses enable the study of a single exogenous environmental shock affecting a single child or all siblings within a family because such analyses are able to exploit variation across families.  

\subsubsection{Exogenous \texorpdfstring{$G$}{} and a PGI based on a regular GWAS}
\label{subsec:lessideal}

In understanding the role of genetic makeup $G_i$, the parameter $\delta$ in \autoref{eq:Kong1} is the main object of interest. However, as \autoref{eq:Kong2} shows, this parameter is biased in a standard GWAS regression of the outcome $Y_i$ on the genotype $G_i$ without controls for parental genotype. In such a GWAS, the coefficient of the child's genotype $G_i$ captures not just $\delta$ but also genetic nurture ($\eta_m$ and $\eta_f$) and assortative mating, sibling effects and ancestry ($w$).

Nevertheless, for the foreseeable future, PGIs based on the results of a between-family GWAS (i.e., one that does not control for parental genotype) will remain significantly more predictive and more readily available than PGIs based on parent--child dyad GWAS summary statistics. Without the possibility to control for parental genotype, PGIs based on standard GWASs will pick up environmental effects due to passive $rGE$ (the darker part of \autoref{fig:diagram}). Initial studies suggest that socioeconomic and cognitive phenotypes are more strongly influenced by familial confounding than are other phenotypes \citep{Trejo2019,selzam2019comparing}. 

Using i) standard GWAS PGIs and ii) controls for parental PGIs in a parent--child dyad dataset or family fixed effects in a sibling sample will yield underestimates of the effect of the child's PGI. As we have seen, the child's PGI based on standard GWASs captures a mixture of direct genetic effects $\delta$, and genetic nurture and other confounding effects, i.e., 
($[\eta_m + \eta_f]/2 + w$) in \autoref{eq:Kong2}. While including parental PGIs account for the component of the child's PGI that captures genetic nurture and other confounding effects ($[\eta_m + \eta_f]/2 + w$), the \textit{parent's} PGIs also pick up \textit{direct genetic} effects $\delta$ (being based on regular GWAS). Thus, controlling for the \textit{parent's} PGIs \textit{overcorrects} the direct genetic effects $\delta$.\footnote{In contrast, in a family-based GWAS, separate PGIs can be constructed for the child's parents deriving from separate SNP coefficients for the child $\delta$ and her parents $\alpha_m$ and $\alpha_f$.} This biases the coefficient of the child's PGI downward as some part of the child's direct genetic effect is now attributed to the parental PGI.

In a family fixed effects specification, the same problem prevails. These designs effectively compare sibling differences in PGIs. A PGI based on a regular GWAS contains direct genetic $\delta$ and genetic nurturing effects and other confounders $[\eta_m + \eta_f]/2 + w$. If one sibling carries more SNPs that, e.g., reflect genetic nurture effects than does the other, the within-sibling PGIs will differ. However, genetic nurture is arguably identical across siblings, reflecting the parental environment shared by siblings. This difference in the estimated PGIs then constitutes measurement error, leading to attenuation bias in the coefficient for $G$ \citep{Trejo2019} and therefore also in the coefficient for $G \times E$.\footnote{Note, however, that if parents respond differently to their children's genetic endowments (see, e.g., \citealt{Sanz-de-galdeano2019}), genetic nurture may actually differ between siblings.}
This also biases the coefficient of the child's PGI downward, as highlighted in the second row of \autoref{tab:Interpretation}.

\subsection{Where we are today} \label{subsec:current}
The ideal $G \times E$ analysis would use GWAS results derived from genetic data on parent--child dyads to construct a PGI. In turn, it would employ the PGI in datasets with genetic data on parent--child dyads and combine this with a source of exogenous variation in the environment, derived from, e.g., natural experiments. The SSGAC and Bristol's Within Family Consortium (WFC) have recently conducted the very first family-based GWASs \citep{Howe2021}. These initial GWASs, however, are not sufficiently powered, as current sample sizes are of the order of 170,000 siblings from a dozen cohorts---substantially smaller than most between-family discovery cohorts, which typically exceed 1 million individuals. Therefore, at least in the near future, researchers might still work with PGIs constructed based on the results of regular GWASs.

\subsubsection{Endogenous \texorpdfstring{$G$}{}}
In the absence of family data, researchers typically include a large number of principal components of the genotypic data as additional controls in the regression to account for population stratification/ancestry. These first (typically 10) principal components of the genotypic data have been shown to be a statistical proxy for respondent ancestry \citep{Price2006} and provide an imperfect but readily available alternative to a within-family analysis. By conditioning on these principal components, the researcher essentially compares individuals within a common lineage and from the same genetic pool.\footnote{For this reason, it is not necessary to include principal components in a within-family analysis.} The substantial reduction in the predictive power of PGIs when family fixed effects are included, compared with analyses that use the principal components of the genetic data, suggests that the use of principal components is not a perfect solution to the omission of parental genotype \citep{Kong2018, selzam2019comparing, koellinger2018using, cheesman2020comparison}. 

The measure of $G$ is thus endogenous in most contemporary $G \times E$ analyses. When $G$ is endogenous, the coefficient of the genetic measure based on standard (between-family) GWASs reflects passive gene--environment correlation $rGE$ due to population stratification, genetic nurture from parents, assortative mating, and sibling and ancestry effects, as illustrated in \autoref{fig:diagram} and \autoref{eq:Kong2}. This case corresponds to the third row in \autoref{tab:Interpretation}. Hence, even when the environment is exogenous (bottom left cell in \autoref{tab:Interpretation}), the measure of $G$ may pick up the effects of parental $G$ and associated environments $E^*$ shaped by the genotypes of parents and other ancestors. As discussed by \cite{Trejo2019}, the conventional (between-family) OLS estimate of the coefficient on $G$ is likely to be biased upward in this case, as the coefficient of the PGI picks up both direct ``child'' genetic effects and environments shaped by parental genes, which typically have the same sign (e.g., child genetic variants associated with higher educational attainment or better health are generally associated with familial environments more conducive to education and health). Therefore, in most cases, the coefficient for the $G \times E$ interaction term will also be biased upward. However, since $G$ in this case reflects the sum of direct genetic effects and passive $rGE$, in theory the interaction between the (exogenous) environment $E$ and these two terms could have opposite signs, leading to a bias in an unknown direction.

\subsubsection{Predetermined \texorpdfstring{$E$}{}}
Commonly analyzed environmental measures such as characteristics of the childhood environment (e.g., area-level unemployment or death rates, distance to facilities, family income) are not shaped by the individual's $G$ but cannot be considered exogenous in $G \times E$ experiments because they are possibly correlated with other environmental characteristics or with parental genotypes. When these measures are analyzed in a family-based sample (i.e., with controls for parental genotype or with family fixed effects), then detecting a statistically significant $G \times E$ interaction coefficient indicates the existence of a ``true'' $G \times E$ effect. The intuition is that by virtue of the family-based nature of the analyses, the measure of $G$ is unbiased and ``randomized'' with respect to the environment. Detecting a $G \times E$ interaction therefore implies the existence of such an interaction rather than a $G \times G$ or $E \times E$ interaction. However, predetermined environmental characteristics tend to cluster together. For example, areas with high unemployment rates tend to have fewer facilities; and family income is strongly associated with parental education and occupation. When $G \times E$ is identified in these cases, the interaction term may actually reflect $G \times E^*$, where $E^*$ is some unobserved correlate of the putative environmental characteristic $E$ (middle column, top row of \autoref{tab:Interpretation}). Thus, the effect of the observed measured environment $E$ may not be causal but rather proxy for the existence of an interaction between $G$ and an unobserved environment $E^*$ that correlates with $E$ and influences the outcome $Y$.

When family-based GWAS results are not available for constructing the PGI but the analyses are performed in family data, the results mirror that in the top middle row except for the downward bias in the $G$ coefficient and therefore in the $G \times E$ coefficient (middle row, middle column of \autoref{tab:Interpretation}). The reason for the downward bias in $G$ and $G \times E$ is identical to that in the case of exogenous $E$ (middle row, first column). 

When the analyst does not have access to family-based datasets, the predetermined environmental characteristic $E$ may additionally reflect parental $G$ (middle column, bottom row of  \autoref{tab:Interpretation}) as a result of familial influences (passive $rGE$). This could, for example, be the case if parental $G$ influences the location of residence or family income $E$. Finally, similarly to the case of exogenous $E$ (first column, bottom row), the coefficient of $G$ is upward biased since the PGI picks up both direct genetic effects and environments shaped by parental genes, which (as discussed before) typically have the same sign. 

\subsubsection{Endogenous \texorpdfstring{$E$}{}}
\label{subsec:endogenousE}
The last column of \autoref{tab:Interpretation} presents the case for endogenous $E$. Here, $E$ is defined as correlated with the error term in the regression model. Endogeneity of $E$ may arise from four sources: reverse causality, omitted variable bias, measurement error, and correlation of the GWAS sample selection with the analyzed environment $E$. The first three sources of endogeneity are common to many econometric analyses \citep[see, among others,][]{Wooldridge2002,angrist2008mostly,Cunningham2021mixtape}, and so we relegate a discussion of these possible biases to \hyperref[appsec:bias]{Appendix~\ref*{appsec:bias}}. Here, we briefly summarize the main ideas and explain endogeneity arising from GWAS sample selection. 

As illustrated in \autoref{tab:Interpretation} (first two rows, last column), even when our measure for $G$ is exogenous, endogeneity of $E$ implies that the coefficient on $G \times E$ cannot be interpreted to represent a causal effect. In fact, it may reflect $G \times E^*$, i.e., a causal effect of some other environment $E^*$ correlated with $E$. Endogeneity of $E$ could also arise from $E$ being shaped by $G$ through active or evocative $rGE$. In this case, $E$ essentially becomes a \underline{mediator} and thereby a ``bad control'' variable (as $E$ is itself an outcome of $G$) in the relationship between $Y$ and $G$.\footnote{This is not the case for a predetermined $E$, since the environment precedes $G$, implying that $E$ is not the result of $G$. A predetermined $E$ therefore rules out active and evocative $rGE$ but not passive $rGE$ (hence our distinction between exogenous, predetermined and endogenous $E$).}  
Furthermore, when parental genotype is not controlled for (bottom row), the environmental effect $E$ may also reflect parental genotype through genetic nurture (see \autoref{fig:diagram}, arrow labeled ``genetic nurture'' from $G_{parent}$ to $E_{child}$).  

The form of endogeneity specific to $G \times E$ analyses occurs when the treatment group in the analysis sample more closely reflects the environmental and demographic characteristics of the GWAS sample used to construct the PGI. In these cases, one may estimate significant $G \times E$ effects without an interaction effect necessarily being present. This is because GWASs---and in turn the PGIs based on the GWAS summary statistics---estimate the genetic effects within the environmental and demographic context of those in the original GWAS sample \citep{domingue2020interactions}. If the treatment group more closely resembles the GWAS sample, $G$ may be more predictive among the treated than among the controls. This difference in predictive power would then be picked up by the $G \times E$ interaction.\footnote{An example may make this clearer: Assume that we are interested in how the effect of being old for one's grade in the school year, captured by being born in September versus August (see \hyperref[sec:application]{Section~\ref*{sec:application}}), interacts with one's PGI for educational attainment in explaining one's educational outcomes. If the original GWAS was performed only on September-born individuals (a very unlikely scenario but useful for illustration), the PGI would likely be more predictive for this group than for August-born individuals since it better captures the environments experienced by September-born students (e.g., older peers) than those experienced by those born in August. A regression of the outcome of interest on $G$, $E$, and $G \times E$ would then lead to a positive estimate on the interaction merely from picking up the additional predictive power among those with September birthdays due to the sample selection in the GWAS without an interaction effect necessarily existing.} Note that this may occur even if $E$ is plausibly exogenous. One way to check whether the environment $E$ is endogenous to the GWAS sample is to test whether $G$ significantly differs across the exogenous environments. This is akin to testing for gene--environment correlation but considering the \textit{exogenous} rather than the \textit{endogenous} environment. Evidence of $rGE$ for \textit{endogenous} environments suggests evocative, active or passive pathways. In contrast, evidence of $rGE$ with \textit{exogenous} environments points to an environment that is possibly endogenous to the GWAS sample selection. The remedy here is to employ alternative GWAS summary statistics to construct PGIs that are not endogenous to the environmental measure in which one is interested. 

\subsubsection{Measurement error}
Although not explicitly included in  \autoref{tab:Interpretation} (it is only mentioned in the table's footnote), a source of endogeneity that deserves special attention is measurement error. Clearly, measures of the environment $E$ could be subject to measurement error, leading to a well-known attenuation bias in the coefficients of $E$ and $G \times E$ \citep{griliches1974errors}. There may also be measurement error in the SNPs used to construct the PGI because most SNPs are imputed rather than measured directly. Measurement error from this source is likely to be small since the imputation quality for common SNPs is high \citep{quick2019sequencing}.\footnote{For imputation quality reasons, rare SNPs (usually defined as those with a minor allele frequency <1\%) are often not analyzed in GWASs and, as a result, not included in PGIs. Excluding these SNPs may decrease the predictive power of PGIs. Still, $\sim$97\% and $\sim$68\% of the genetic variation for common and rare variants, respectively, is captured by imputation; see \cite{Yang2015}.} 
However, even when the SNPs are measured without error, since the discovery GWAS samples are not infinitely large, there is measurement error in the estimated GWAS coefficients and hence in the PGIs. The constructed PGI is therefore a noisy proxy of the ``true'' PGI (see \hyperref[sec:polygenicindices]{Section~\ref*{sec:polygenicindices}}). 

Simply ignoring measurement error $G$ is not a good strategy. As explained in \hyperref[subsec:current]{Section~\ref*{subsec:current}}, passive $rGE$ is likely to bias the coefficient of $G$ upward, while classical measurement error will attenuate the identified effect of $G$ and $G \times E$. Hence, these two sources of bias lead to an overall bias of unknown direction. A promising way to address attenuation bias of the PGI is to apply instrumental variables (IVs), as first suggested by \citet{DiPrete2018}. Since there are often multiple discovery cohorts available or one can split the GWAS discovery sample in two, it is relatively straightforward to construct two ``independent'' PGIs in a prediction sample and use these as instruments for one another. The most efficient way to combine this information is to use the recently proposed ``obviously related instrumental variable'' method \citep{gillen2019experimenting,VanKippersluis2020}.\footnote{An alternative is to use a Structural Equation Model (SEM) with the two independent PGIs treated as measurements of a latent underlying factor \citep{Tucker-Drob2017}.} 

\subsubsection{Summary}
Our discussion of the scenarios in \autoref{tab:Interpretation} makes clear that one can keep bias in $G \times E$ analyses manageable when exploiting exogenous or predetermined environments rather than endogenous environments. Exogenous variation in the environment $E$ can be analyzed with the usual toolkit available to the applied econometrician: randomized controlled trials, difference-in-differences methods, regression discontinuity designs, instrumental variables, or other (quasi-)experimental methods \citep{Schmitz2017}. In this way, the environmental measure is independent of $G$, and one can draw causal conclusions from the environmental exposure. With within-family GWAS samples currently too small to construct PGIs based on their results, regular (between-family) GWAS results are likely to be used to construct PGIs for $G \times E$ analyses. The coefficient of $G$ in such analyses will be biased either upward or downward depending on whether family- or population-based analysis samples are used. Measurement error in the PGI will also attenuate the coefficients of $G$ and $G \times E$ in regression models, but this can be partly overcome by IV approaches.

\subsection{The functional form} \label{sec:missp}
Even when one considers exogenous measures of $G$ and $E$, a threat to an unbiased estimation of $G \times E$ is misspecification of the functional form. Which functional form to choose depends on how $G$ and $E$ are measured and whether their putative interplay is thought to be additive, multiplicative, or more complex. A careful study of one's data is therefore an essential prerequisite for any empirical tests.

As discussed in \hyperref[sec:G]{Section~\ref*{sec:G}}, $G$ is typically measured by a PGI, a continuous variable. In line with the idea of considering $G$ to be the genetic propensity to developing a certain trait, treating $G$ as a continuous variable is a natural starting point. The relevant PGI is typically pinned down by the choice of the outcome of interest. For example, if one studies educational attainment, a natural choice for $G$ is the PGI for education. However, this need not always be the case. Any PGI could be used if warranted by theory or for empirical reasons. For example, the PGI for educational attainment has been shown to predict a variety of outcomes, such as social mobility and wealth \citep{Belsky2016,Belsky2018mobility,papageorge2020genes, barth2020genetic}.

The choice of the environmental component $E$ leaves many degrees of freedom. To guide the empirical researcher interested in estimating $G \times E$, we therefore advocate for the use of theory and/or a clear source of exogenous variation in $E$ to select from the possible measures of environment $E$. For example, biological or social science theories may stipulate a clear hypothesis of the most relevant measure of environment $E$ to be selected, such as maternal smoking invoking epigenetic expression or social norms restricting risky behaviors such as smoking and alcohol consumption \citep{cho2015alcohol}. Alternatively, the starting point could be an exogenous source of variation in the environment $E$. For example, quasi-natural experimental designs can be used to isolate variation in environmental exposure to avoid potential confounders  \citep[see, e.g.,][]{Schmitz2017,Barcellos2018}. The advantages of focusing on exogenous changes in the environment are that it enhances the scientific validity of the results and facilitates interpretation of the estimated effect, as illustrated in \autoref{tab:Interpretation}. A careful operationalization of the environment $E$ allows a causal interpretation of the direct effect of the environment and a better understanding of whether and how such an effect varies with an individual's genetic predisposition $G$. A downside is that not every theoretical hypothesis easily lends itself to causal identification and not every social science dataset containing genetic information can be easily linked with exogenous variation in the environment. 

Measures of $E$ can be binary, categorical, or continuous. Whereas continuous environmental exposures may enter linearly, we would recommend first dividing the environmental exposure into discrete categories (or bins) to study the most appropriate functional form. When the relationship is nonlinear, entering $E$ linearly may lead to serious biases \citep{hainmueller2019much}. For simplicity, we here consider the case where $E$ is a binary variable, i.e., there is a ``treatment'' and a ``control'' group.\footnote{A continuous variable can always be dichotomized into two values, such as high and low exposure to $E$. More importantly, this setup corresponds to the textbook case where the environment of interest is a treatment or a policy change that can be estimated via the potential outcomes framework \citep{Neyman1923,Fisher1935,Rubin1974}.} We can then separately plot the relationship between $G$ and the outcomes for the treated and control groups. Any difference in the $Y$-$G$ relationship between the treated and control groups is evidence in favor of $G \times E$. Differences in levels suggest that including $G$ and $E$ separately should suffice; differences in slopes between the two groups suggest a simple additive and linear specification with an interaction term; differences in only some parts of the distribution of the PGI suggest a more complex form of $G \times E$ interplay.\footnote{For a more thorough discussion of nonparametric analyses of gene--environment interplay in the context of a structural equation model (SEM), see \cite{Briley2015}. For more flexible functional forms additionally modeling possible heteroskedasticity, see \cite{Domingue2020.09.08.287888}.} These plots could potentially also detect whether the outcome is better specified as linear or logarithmic, although this choice should also be justified on more fundamental grounds, as a logarithmic specification renders the interaction term multiplicative instead of additive and may be less relevant from a policy perspective \citep{vanderweele2014tutorial}. 

The nonparametric plot of the relationship between $Y$ and $G$, separately for the treated and control group, will also shed light on whether to include nonlinear terms for $G$ (and $E$ when it is not dichotomized). As described in \autoref{sec:econModel}, it is generally advisable to include quadratic terms for $G$ and $E$ in the model to properly account for behavioral responses to a certain genetic or environmental endowment. When exogeneity of $G$ or $E$ can only be guaranteed conditional on inclusion of certain control variables, one should always add interaction terms between each control variable and both the $G$ and $E$ terms. These additional interaction terms, $\mu_{g}$ and $\mu_{e}$ in \autoref{eq:function}, aim to capture any residual correlation between the controls and either genes or environment and are essential for estimating an unbiased effect of the interaction term \citep{Keller2014}. 

\subsection{Checklist for the applied researcher} \label{sec:checklist}
To conclude this section, we provide a simple checklist for applied researchers interested in estimating a form of gene--environment interplay. We assume that by the time the researcher is going through the checklist, she has already made a decision regarding the construction of the variables $Y$, $G$, and $E$:
\begin{enumerate}
    \item \textbf{Perform power calculations:}  
    Since the anticipated effect sizes of interaction terms are typically an order of magnitude smaller than those of the main effects, it is strongly advised to conduct ex ante power calculations before any empirical analysis is done. This avoids underpowered statistical tests and reduces the risk of finding false positives and negatives. Arguably the best way to mimic the actual statistical power of an empirical test for the interaction term is through simulations \citep{Duncan2011}. These allow one to specify the exact model that one intends to estimate and can account for possible adjustments to the standard error. 
    \item \textbf{Check for gene--environment correlations $rGE$}:
    Check the extent to which genes correlate with the environment of interest. Finding no significant evidence for $rGE$ would help support the assumption that a truly exogenous environment is exploited in the analyses.
    \item \textbf{Choose the correct functional form:} 
    Although $G \times E$ analyses are often conducted using linear interaction terms, it is advised to use descriptive analyses to check for nonlinearities in the relationship. To be as nonparametric as possible, one could plot the relationship between $G$ and $Y$ separately for the treated and control groups. 
    \item \textbf{Perform $G \times E$ analysis:} 
    Perform the $G \times E$ analyses using the functional form determined in the previous step. Include all relevant control variables (e.g., principal components when within-family designs are not used). Interpret the findings in view of the mechanisms described in \hyperref[sec:econModel]{Section~\ref*{sec:econModel}} and assess the bias in the estimate based on \autoref{tab:Interpretation}.
    \item \textbf{Correct inferences for heteroskedasticity and multiple hypothesis testing:} As explained in \hyperref[sec:econModel]{Section~\ref*{sec:econModel}}, behavioral responses to $G$ and $E$ almost by definition invoke heteroskedasticity in the error term in $G \times E$ specifications. It is therefore prudent to at least use robust standard errors in the analysis, and the analyst may wish to conduct heteroskedasticity tests as in \citet{Domingue2020.09.08.287888} if the variance of the outcome is expected to vary systematically with the environmental exposure. If the analysis involves many different outcomes explored under a more data-driven approach, one may want to correct for multiple hypothesis testing. 
    \item \textbf{Perform robustness checks:} Particularly when there is still doubt about the exogeneity of $G$ or $E$, one could use robustness checks to test whether the estimate of interest represents a true $G \times E$ effect. For example, one could use placebo tests (e.g., use unrelated PGIs or random permutations of the PGI; see \cite{Muslimova2020b}). One could also try to replicate the result in a different setting or dataset to investigate to what extent it is generalizable to other settings. 
\end{enumerate}

\section{An empirical illustration of \texorpdfstring{$G \times E$}{} interplay}
\label{sec:application}

To provide a concrete illustration of the step-by-step procedure that can be followed, we present a simple example of how to estimate a gene--environment interplay model. In this illustrative application, we focus on the difference in school test scores between younger and older pupils in the same grade and estimate whether this effect is moderated by the child's genetic predisposition for educational attainment. In other words, we use the pupil's month of birth as a measure of the environment ($E$), the PGI for educational attainment as a measure of her genetic predisposition ($G$), and explore their joint influence on school test scores ($Y$).

We leverage the fact that the month of birth determines \textit{when} pupils start school in England. More specifically, pupils start school in the school year (i.e., counting from September) they turn five. This means that at one extreme, children born on August 31 start their primary schooling when they are four years and one day old, whereas those who are born one day later, on September 1, start primary school on their fifth birthday. Hence, the latter group is a full year (minus one day) older than students born on August 31. At age four, this is 25\% of their lifetime---a non-negligible difference. 

We focus on this quasi-exogenous change in the environment because, on average, older pupils have been shown to perform better on educational tests than their younger counterparts in the same grade \citep[see e.g.,][]{Bedard2006,Fredriksson2005,Black2011,Crawford2010,Ritchie2018}, and the evidence suggests this has significant long-term implications \citep{Page2019}. \cite{Crawford2010} find that the main driver for the performance gap between younger and older pupils is an absolute age effect: those born earlier in the school year are up to one full year older at the time that they take the exam than their younger peers. We explore whether a high PGI for educational attainment exacerbates test score differences by month of birth or attenuates these absolute age effects, as well as how the effects change as children age.

We investigate the month-of-birth effect using the Avon Longitudinal Study of Parents and Children (ALSPAC). ALSPAC is a cohort study in which pregnant women living in Avon (UK) who had an expected delivery date between 1 April 1991 and 31 December 1992 were invited to take part. The initial number of pregnancies was 14,541 and number of fetuses 14,676. Among these, 13,988 children were alive at age 1.\footnote{The sample size for analyses using data collected after the age of seven is 15,454 pregnancies, resulting in 15,589 foetuses. Of these 14,901 were alive at 1 year of age. For more information on ALSPAC, see \cite{Boyd2013} and \cite{Fraser2013}. Please note that the study website contains details of all the data that is available through a fully searchable data dictionary and variable \href{http://www.bristol.ac.uk/alspac/researchers/our-data/}{search tool} at http://www.bristol.ac.uk/alspac/researchers/our-data/.} 
We use ALSPAC for three main reasons. First, the cohorts covered have to abide by strict rules on when children start school relative to their date of birth.\footnote{Although the government currently allows some degree of flexibility, whereby parents can choose to send their child to school a year earlier or hold them back a year, this was not possible at the time that pupils in our dataset started their primary schooling. Note also that children do not repeat school years in England.} Hence, there is a strong discontinuity in children's starting age between those born just before and just after the threshold of September 1. Although births are often planned, implying that month of birth is a choice variable and therefore endogenous, we find no evidence that being born just before or after the end of August is systematically related to children's or parental background characteristics; we show this below. 
Second, the majority of cohort members have been genotyped. We create a PGI for educational attainment by meta-analysing GWAS summary statistics from the UK Biobank and 23andMe and correcting for linkage disequilibrium between SNPs with the software package LDpred \citep{Vilhjalmsson2015}.\footnote{The polygenic index is constructed using LDpred version 1.0.5, and Python, version 3.6.6. LDpred is a software package based on Python that adjusts the GWAS weights for LD using a Bayesian approach. We re-weight the SNP effects on the basis of LD and the supposed fraction of causal SNPs, which we set to 1, as is standard practice for behavioral traits \citep{cesarini2017genetics}. The polygenic index includes all SNPs, that is 1,177,817 SNPs after filtering for HapMap3 SNPs at the coordination step.} We standardize the PGI to have mean 0 and standard deviation 1 in the analysis sample. Third, ALSPAC contains an extremely rich set of child outcomes, including those from administrative sources. Specifically, we use children's performance on exams taken at five time points. We use an entry assessment test, taken by all pupils about to start primary school (at age 4), and four nationally set examinations taken at ages 7, 11, 14 and 16 (also known as Key Stage 1 (KS1), 2 (KS2), 3 (KS3) and 4 (KS4 or GCSE) examinations, respectively). Children's scores are obtained from the National Pupil Database, a census of all pupils in England within the state school system, which is matched to ALSPAC.\footnote{At age 18, study children were sent `fair processing' materials describing ALSPAC’s intended use of their health and administrative records and were given clear means to consent or object via a written form. Data were not extracted for participants who objected, or who were not sent fair processing materials.} For each of the tests, we use an average score for the child's mandatory subjects.\footnote{For KS1, this is an average of the child's reading, writing, spelling and mathematics scores; KS2 includes reading, writing, science and mathematics tests. For KS3 and KS4, the final score is an average of the child's English, mathematics and science scores.} In our analyses, all test scores have been standardized to have mean 0 and standard deviation 1 in the analysis sample.

Although there are many advantages of using ALSPAC, as with any data, there are also downsides. The main disadvantage is that there is no systematic data collection on siblings or fathers.\footnote{Although today ALSPAC focuses on data collection from other family members such as fathers and siblings, this was not the case when the survey was set up. Hence, we cannot exploit this information.} This implies that we cannot perform a within-family analysis and hold family background constant, which would help alleviate some concerns about the potential endogeneity of month of birth and the endogenous nature of $G$. Hence, as detailed in \autoref{tab:Interpretation}, the interpretation of the $G$ and $G \times E$ coefficients is more complicated. We return to this below.

\subsection{Identification strategy} \label{sec:identification}
We are interested in the difference in test scores between the oldest and youngest children in the school year, and we structure our analysis following the checklist suggested in \hyperref[sec:checklist]{Section~\ref*{sec:checklist}}. To identify the effect of interest, we use a regression discontinuity design (RDD), specifying the treated and control groups as pupils born after and before September 1, respectively. To empirically check the validity of our identification strategy, we begin by exploring the raw data, plotting the trends in educational attainment by month of birth and examining the correlations between treatment, educational attainment, and the PGI.

\begin{figure}[ht]
\centering 
\includegraphics[width=0.85\linewidth]{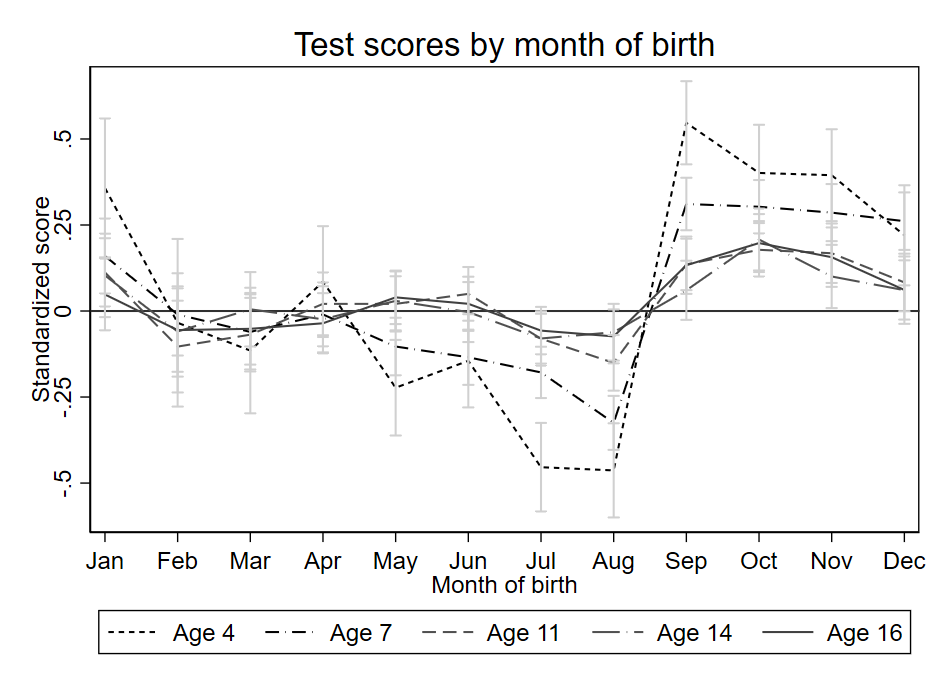}
\caption{Standardized test scores at different ages by month of birth.}
\label{fig:MoB}
\end{figure}

\autoref{fig:MoB} presents the standardized test scores on the vertical axis by pupils' month of birth on the horizontal axis. We show the average test scores by month of birth for the five exam results observed in ALSPAC, taken at ages 4, 7, 11, 14 and 16. The figure shows a clear discontinuity in test scores for those born in September and beyond in comparison to the scores of pupils born before September, with the latter group performing significantly worse on all five tests observed. This discontinuity is largest for the age-4 test, for which the (proportional) difference in age between those born in August and September is largest (up to 25\%): those born in September perform approximately one standard deviation better than their August-born peers. This difference reduces as children age (i.e., as the proportional difference in age between the oldest and youngest in the year decreases), but it remains sizable and statistically significant across all assessments.

Our main interest is in the discontinuity in test scores at the cutoff and in investigating whether (and how) this varies with an individual's genetic predisposition for educational attainment. Thus, we restrict the sample to those born between June and November: three months before and three months after the threshold.\footnote{The bandwidth of three months has been chosen to balance the need for efficiency (and the associated need to use as many data points as possible) and the danger of bias (and the associated need to remain as close to the cutoff as possible). Our results are robust to the use of different bandwidths.} We focus on the age-4 test in our subsequent descriptive statistics and analysis, though we show the robustness of our findings to the use of the later Key Stage tests below. 

To explore whether the month of birth can be considered as good as random, \autoref{tab:descr} reports descriptive statistics for a set of pupil and family characteristics by treatment status. We include only covariates observed before (or at) birth, as any variables measured later in life could be directly or indirectly affected by the treatment. Although \autoref{tab:descr} shows some significant differences, they are very small, and there is no strong evidence that the treated group is systematically different from the control group.\footnote{\citet{Buckles2013} and \citet{Schwandt2017} find differences in maternal characteristics for births throughout the year, but their focus is on documenting differences across seasons rather than selection of births just before or after a precise cutoff. Their results are therefore not in contrast to ours: long-term seasonality trends do not jump at a specific birth date.} For example, mothers in the treated group are slightly less likely to have O-levels but more likely to be in the semi-skilled social class. However, both differences are small and only significant at the 10\% level. We find no other differences between the treated and control groups with respect to maternal age at first birth, smoking and mental health during pregnancy, marital status, partner's education, and the child's birth weight. Finally, there are no differences by treatment status in the child's or mother's PGI for educational attainment. This provides at least suggestive evidence that individuals' month of birth can be considered unrelated to the child and family background characteristics that we observe here. There is also no evidence of $rGE$.

\begin{table}[H]
\caption{Descriptive statistics of pupil and family characteristics by treatment status.}
\centering
\begin{tabular}{lccccccccccc}
\toprule
&\multicolumn{3}{c}{Treated} &\multicolumn{3}{c}{Control}           &\multicolumn{3}{c}{\textit{t} test}           \\
& \textit{N} & Mean & & \textit{N} & Mean & & \textit{p} value \\
\midrule
Mother's age at first pregnancy (in years)&        2062&      25.138&            &        2168&      25.257&            &        0.431\\
Mother smoked cigarettes during pregnancy&        1927&       0.167&            &        2052&       0.168&           &       0.896\\
Mother's anxiety score during pregnancy&        1888&       4.651&            &        2037&       4.659&            &       0.946\\
Mother's depression score during pregnancy&        1887&       4.245&            &        2038&       4.211&            &       0.714\\
Mother is married   &        2061&       0.843&            &        2169&       0.859&            &        0.153\\
Mother's education: Vocational&        2047&       0.096&            &        2146&       0.088&            &        0.361\\
Mother's education: O-level&        2047&       0.338&            &        2146&       0.366&            &        0.056\\
Mother's education: A-level&        2047&       0.248&            &        2146&       0.241&            &        0.609\\
Mother's education: Degree&        2047&       0.157&            &        2146&       0.156&            &        0.882\\
Partner's education: Vocational&        1976&       0.082&            &        2085&       0.071&            &        0.188\\
Partner's education: O-level&        1976&       0.197&            &        2085&       0.218&            &        0.101\\
Partner's education: A-level&        1976&       0.275&            &        2085&       0.278&            &        0.836\\
Partner's education: Degree&        1976&       0.214&            &        2085&       0.228&            &        0.258\\
Mother's social class: II&        1713&       0.325&            &        1864&       0.326&            &        0.948\\
Mother's social class: III (Non-manual)&        1713&       0.422&            &        1864&       0.434&            &        0.471\\
Mother's social class: III (Manual)&        1713&       0.067&            &        1864&       0.072&            &        0.576\\
Mother's social class: IV&        1713&       0.101&            &        1864&       0.083&            &        0.058\\
Mother's social class: V&        1713&       0.016&            &        1864&       0.014&            &        0.652\\
Child's birthweight (in grams)&        2089&    3448&            &        2189&    3451&            &        0.875\\
Child's PGI for educational attainment&        2114&       0.016&            &        2209&       0.037&            &        0.499\\
Mother's PGI for educational attainment&       1526&       0.036&            &        1533&       0.022&            &        0.688\\

\bottomrule
\addlinespace[.75ex]
\end{tabular}
\label{tab:descr}
\caption*{\footnotesize \noindent \textit{Notes:} The table shows the sample size and means for a set of child and family characteristics observed before or at birth. Columns (1) and (2) show this for the treated group; columns (3) and (4) show this for the control group. Column (5) shows the \textit{p} value from a \textit{t} test of whether the means for the two groups are significantly different from one another. We report descriptive statistics for the maximum sample; the sample sizes in the estimation below depend on the missingness in the outcome variable.}
\end{table}

\subsection{Power calculations} 
\label{sec:powerCalc}
We use simulations to perform \textit{ex ante} power calculations to estimate the minimum detectable effect (MDE) for the $G \times E$ coefficient in our setting. Since the interaction term captures a form of nonlinearity in the relationship between $G$, $E$ and the outcome $Y$, simulations are arguably the most flexible and reliable way to approximate the actual statistical power of our empirical test \citep{Duncan2011}. We simulate the following (simplified) $G \times E$ model 1,000 times and approximate statistical power using the share of simulations with a $p$ value $<0.05$:\footnote{The STATA code for these simulations and a simplified example code for simulating the statistical power of the empirical specification is available at our \href{http://github.com/geighei/GxE_4practitioners}{GitHub} repository. The example code allows one to specify a binary or continuous environment, different sample sizes, and other expected MDEs.}
\begin{equation} \label{eq:power}
    Y_i = \alpha + \beta_G G + \beta_E E + \beta_{\times} (G \times E) + \varepsilon.
\end{equation}
We set the parameters based on empirical results in the literature. Since our outcome variable $Y_i$ is always standardized to have mean 0 and standard deviation 1, we assume that $\alpha=0$ and that the error term is drawn from a standard normal distribution $\varepsilon \sim N(0,1)$. We also assume that $G$ is standard-normally distributed ($G \sim N(0,1)$), which is a good approximation of the distribution of the standardized PGI, and following \cite{Allegrini2019}, we set $\beta_G=0.259$.
Since approximately half of our sample is treated (i.e., born between September and November), we randomly assign 50\% of the simulated observations to have $E=1$ and the other half $E=0$. Following \cite{Crawford2010}, we assume that $\beta_E=0.90, 0.60, 0.35, 0.20$ and $0.13$ when the outcome is the age-4 (entry assessment) score, KS1, KS2, KS3, and KS4, respectively.\footnote{Note that \cite{Crawford2010} do not report estimates for the age-4 entry assessment test. We hypothesize this to be $0.90$, in line with the pattern of greater effect sizes for younger children. Indeed, as we show below, our results for the Key Stages are remarkably similar to these authors' estimates: 0.71 (KS1), 0.39 (KS2), 0.23 (KS3) and 0.28 (KS4), as shown in \autoref{tab:MoB_ks}.} Finally, given our ALSPAC data, we set the sample size in the power calculations to $N=1,000$ for the entry assessment score, $N=3,500$ for Key Stages 1, 2 and 4, and $N=3,000$ for Key Stage 3.

\begin{figure}[H]
\centering 
\includegraphics[width=0.7\linewidth]{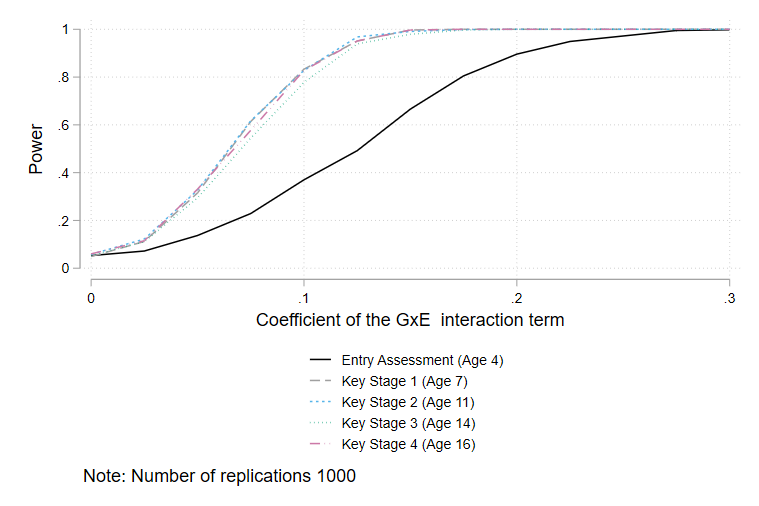}
\caption{Power calculations.}
\label{fig:power}
\end{figure}

\autoref{fig:power} shows the statistical power that we can expect for each outcome variable and for different magnitudes of the interaction coefficient $\beta_{\times}$ from \autoref{eq:power}. Following the literature,\footnote{\cite{Crawford2010} do not report heterogeneity of treatment effects based on either individual or parental characteristics. However, \cite{Black2011} report heterogeneity of effects based on a composite measure of parental background (see Table 7 in their paper). Moving from the bottom 25\% to the top 75\% of the distribution (a shift equivalent to about 2 standard deviations in our PGI), the effect of early age at school entry more than doubles for outcomes such as education, earnings at age 35, and teenage pregnancy but does not differ for outcomes such as IQ or mental health. We take a conservative approach and consider a change of about one-quarter of the main effect of the environment.} %
we expect the magnitude of the interaction effect to be about one-quarter of the size of the main effect of the environment. Therefore, we expect $\beta_{\times}=0.225, 0.15, 0.0875, 0.05$ and $0.0325$ when the outcome is the entry assessment score, KS1, KS2, KS3, and KS4, respectively. Hence, we have more than 90\% power to estimate the interaction effect for the entry assessment score and about 75\% power for KS1, although we are underpowered for the other outcomes.
More generally, we are well powered ($>80\%$) to estimate an interaction coefficient greater than 0.1 for the Key Stage outcomes and greater than 0.175 for the entry assessment. 

\subsection{Gene--environment correlation \textit{rGE}.} \label{sec:rGE}
To more formally check for the existence of gene--environment correlation, we explore whether the treated and control groups have systematically different PGIs, which would suggest selection into treatment based on genetic characteristics.

\begin{figure}[H]
\centering 
\includegraphics[width=0.49\linewidth]{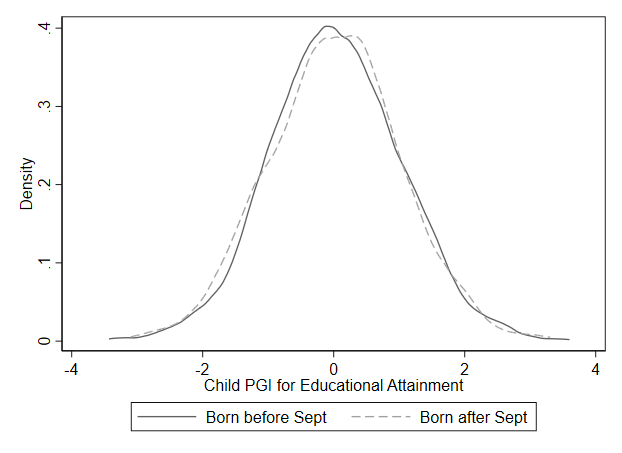}
\includegraphics[width=0.49\linewidth]{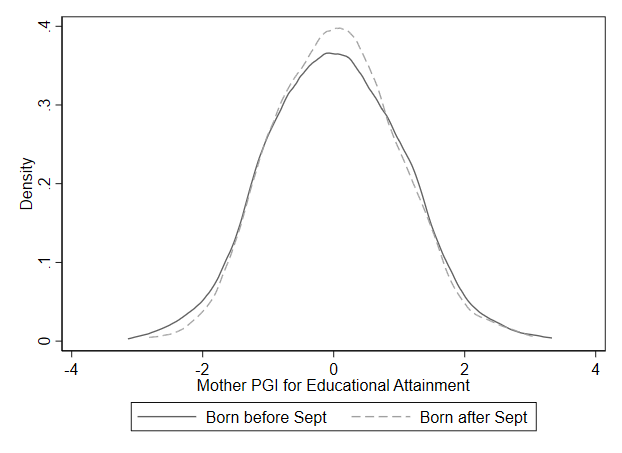}
\caption{Densities of children's and mothers' PGIs by treatment status.}
\label{fig:kdens_treat_PGI}
\end{figure}

The left-hand panel of \autoref{fig:kdens_treat_PGI} plots the density of the child's PGI for the treatment and control groups, showing little difference in their distributions. The right-hand panel presents the PGI for the children's mothers by treatment and control group, showing similar overlapping distributions. 
The polychoric correlations between the treatment indicator and the child's PGI ($\rho = -0.013$, s.e. 0.019) or the maternal PGI ($\rho = 0.009$, s.e. 0.023) are both very small and statistically indistinguishable from zero. This suggests there is no gene--environment correlation. 

\subsection{Predictive power of the PGI} \label{sec:PGI}
We next explore the predictive power of the PGI for children's educational attainment at different ages.  \autoref{tab:PredictivePower} shows the estimates from OLS regressions of the outcome on the child's PGI for educational attainment. The five columns represent the five test scores of interest, taken at ages 4, 7, 11, 14 and 16. The results show that each one-standard-deviation increase in the PGI is associated with an increase in test scores by 0.16 standard deviations at age 4, 0.26 standard deviations at age 7 (KS1), and between 0.32 and 0.36 standard deviations at ages 11--16, suggesting that the predictive power of the PGI increases as the child ages. 

\begin{table}[H]
\caption{OLS estimates of the effect of the PGI for EA on test score at different ages.}
\centering
{\footnotesize
\begin{tabular}{lcccccccccccccccccccc}
\toprule
            &\multicolumn{1}{c}{Entry Assessment}&\multicolumn{1}{c}{Key Stage 1}&\multicolumn{1}{c}{Key Stage 2}&\multicolumn{1}{c}{Key Stage 3}&\multicolumn{1}{c}{Key Stage 4}\\\cmidrule(lr){2-2}\cmidrule(lr){3-3}\cmidrule(lr){4-4}\cmidrule(lr){5-5}\cmidrule(lr){6-6}
\midrule
PGI for EA         &       0.163\sym{***}&       0.255\sym{***}&       0.344\sym{***}&       0.357\sym{***}&       0.318\sym{***}\\
            &     (0.028)         &     (0.015)         &     (0.015)         &     (0.017)         &     (0.016)         \\
\midrule
$R^2$          &       0.085         &       0.096         &       0.128         &       0.134         &       0.128         \\
Observations&        1094         &        3436         &        3610         &        3073         &        3579         \\

\bottomrule
\addlinespace[.75ex]
\end{tabular}
\label{tab:PredictivePower}
}
\caption*{\footnotesize \noindent \textit{Notes:} The test score and the PGI (PGI) for educational attainment (EA) are standardized to have mean 0 and standard deviation 1 in the analysis sample. All regressions control for gender and the first ten principal components of the genetic data. Robust standard errors in parentheses. * $p < 0.10$, ** $p < 0.05$, *** $p < 0.01$.}
\end{table}

\subsection{Functional form} \label{sec:form}
\begin{figure}[H]
\centering 
\includegraphics[width=0.8\linewidth]{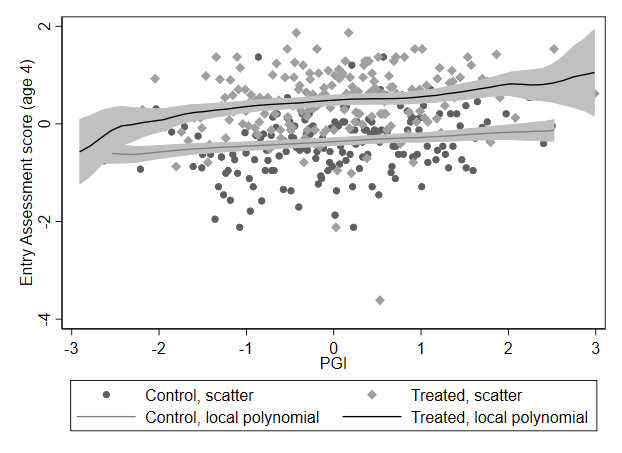}
\caption{The relation between the PGI for educational attainment and the entry assessment score (age 4) by treatment and control group. In this figure, the distribution of the PGI is trimmed to be between -3 and +3 to avoid nonlinear overfitting of outliers.}
\label{fig:PGIxTreat_ea}
\end{figure}

As mentioned above, we focus on the age 4 entry assessment test, which showed the largest treatment effect \autoref{fig:MoB}. \autoref{fig:PGIxTreat_ea} plots the relationship between the PGI and the entry assessment score in a non-parametric fashion separately for the treatment and control group (i.e. comparing children born in September, October and November with those born in June, July and August). The figure shows that both relationships are positive, with some suggestion that the slope is slightly steeper for the treatment group. Furthermore, the figure suggests that we can approximate the relationship between the PGIs and the outcome as linear, with both lines of best fit being approximately linear between -2 and +2 standard deviations of the PGI (i.e. for at least 95 percent of the data). 

\subsection{Empirical specification} \label{sec:analysis}
We are interested in the discontinuity in test scores by month of birth, with the difference between those born before and after September constituting the treatment. To estimate the main effect of the PGI and the environment (i.e., born in or after September), we follow a standard regression discontinuity design specification:
\begin{equation}\label{eq:E}
\begin{aligned}
    TestScore_i = \beta_0 + \beta_{G} G_i + & \beta_{E} E_i + \\
        & \delta_a MoB_i + \delta_b (MoB_{i} \times E_i) + \\ 
        & \delta_c male_i + \delta_d (Male_i \times E_i) + \\
        & \delta_{e} YoB92_i + \delta_{f} (YoB92_i \times E_i) + \\
        & \sum_{p=1}^{10} \gamma_{p} PC^p_i + \sum_{p=1}^{10} \delta_{p} (PC^p_i \times E_i ) + u_i,
\end{aligned}
\end{equation}
where $TestScore_i$ is the child's test score performance and $E_i$ is the environment of interest: a dummy that equals one for treated individuals (i.e., those born in the first three months of the academic year) and zero for the controls (i.e., those born between June and August); $MoB_i$ is the running variable, capturing the trend in $TestScore_i$ by month of birth for those born before September (note that this variable runs from -3 to 2, capturing the calendar months June--November, with September set to 0). The coefficient on $MoB_i \times E_i $ captures any changes in slope for those born after September. We control for gender (denoted by $Male_i$), a dummy for birth in 1992 ($YoB92_i$; capturing potential differences in test scores between the two ALSPAC birth cohorts), individuals' PGI ($G_i$), the first ten principal components of the genetic data ($\sum_{p=1}^{10}PC^p_i$), and interactions between all control variables and the environment. We compare this to the specification that incorporates gene--environment interactions, given by 

\begin{equation}\label{eq:GxE}
\begin{aligned}
        TestScore_i = \delta_0 + & \delta_{G} G_i + \delta_{E} E_i + \delta_{\times} (G_i \times E_i) + \\ 
        & \delta_1 MoB_i + \delta_2 (MoB_i \times G_i) + \delta_3 (MoB_i \times E_i) + \delta_4 (MoB_i \times G_i \times E_i) + \\ 
        & f_1\left(X_i,G_i,E_i\right)
        + f_2\left(\sum_{p=1}^{10} PC^p_i,G_i,E_i\right) + e_i.
\end{aligned}
\end{equation}

\autoref{eq:GxE} includes the two main effects---the PGI ($G_i$) and treatment status ($E_i$)---and the running variable $MoB_i$: a linear specification of individuals' month of birth. We include interactions between these main effects and all (demeaned) controls $X_i$, as suggested in \cite{Keller2014}.\footnote{$f_1\left(X_i,G_i,E_i\right)$ denotes all interactions between the covariates (i.e., gender and year of birth) and $G_i$, as well as between the covariates and $E_i$; $f_2\left(\sum_{p=1}^{10} PC^p_i,G_i,E_i\right)$ is a short-cut for all interactions between the ten (demeaned) principal components and $G_i$ and between the ten (demeaned) principal components and $E_i$. It is important to demean all covariates before interacting them to facilitate interpretation of the coefficients. For example, consider the relationship between an outcome $Y$ and two variables, one continuous (but demeaned) $Z$, denoted as $\tilde{Z}=Z-\frac{1}{n}\sum{Z}=Z-\bar{Z}$, and one dichotomous $W$ and their (demeaned) interaction $\tilde{Z} \times W$. The interpretation of $\beta_{\times}$ in the linear equation $Y=\alpha+\beta_{Z}\tilde{Z}+\beta_{W}W+\beta_{\times}(\tilde{Z} \times W) + u$ is the average difference in $Y$ between $W=1$ and $W=0$, evaluated at the sample average $Z=\bar{Z}$. Failing to demean $Z$ implies that $\beta_{\times}$ captures the average difference in $Y$ evaluated at $Z=0$, which might not be plausible or may be outside the support of the data. We demean all control variables (other than $E$) in \autoref{eq:E} and \autoref{eq:GxE}.
}
Hence, the coefficient for $G_i$, $\delta_G$, captures the change in test scores associated with a one-standard-deviation increase in the PGI for the control group ($E=0$) with average characteristics $X_i$, while $\delta_E$ estimates the treatment effect for pupils with an average PGI of 0 and average characterstics $X_i$. Finally, $\delta_{\times}$ is our estimate of interest, capturing whether the discontinuity in test scores differs by individuals' PGI. 

We showed earlier (\autoref{tab:descr}), that our environment, i.e., being older in one's grade, is as good as random. This suggests, based on \autoref{tab:Interpretation}, that $\delta_E$ is unbiased, capturing the effect of being older in one's grade on one's test scores. In contrast, as we do not observe siblings' or both parents' genotypes in ALSPAC and therefore cannot control for family genotypes, the PGI potentially captures spurious correlation with the parental and family environment. Indeed, in addition to capturing one's genetic predisposition, the coefficient on $G$ may reflect other environmental factors due to genetic nurture or passive $rGE$ via the parental genotype. Since the genetic effect and genetic nurture typically have the same sign, $\delta_G$ is likely to be biased upward. This in turn implies that the coefficient on the $G \times E$ interaction term, $\delta_{\times}$, may additionally capture the interplay between different environmental components via genetic nurture, i.e., $E \times E^{(*)}$. It is important to take this into account when interpreting the findings.

\subsection{Results} \label{sec:results}
We quantify the main effects and their interactions for the analysis of the entry assessment score in \autoref{tab:MoB_ea}.\footnote{As our analysis is hypothesis driven, testing one specific PGI and one specific environment with the same outcome measured at different ages, we do not account for multiple hypotheses testing here.} Columns (1) and (2) present the results from estimating \autoref{eq:E} and \autoref{eq:GxE}, respectively. For both specifications, we find that treated students are scoring on average just over one standard deviation higher on their educational test than the controls. Compare this with \autoref{fig:MoB} where treatment (roughly the change from August to September) is about 1 standard deviation (jumping from $\sim$ -0.5 to $\sim$ +0.5 standard deviation). Furthermore, for each month that one is born later (MoB), the test score reduces by 0.15 of a standard deviation for the controls (MoB); this trend is slightly less pronounced (though insignificantly so) for the treated. The coefficient on PGI in column (1) suggests that a one-standard-deviation increase in the PGI is associated with an increase of 0.155 standard deviations in the entry assessment score---a result similar to the estimate of 0.163 in \autoref{tab:PredictivePower}.\footnote{In the specification that includes the interaction term (column 2), the main effect of the PGI, estimating the counterfactual effect of the PGI for those born in September (i.e., MoB=0) but who are not treated (Treated=0), is no longer significant. Instead, the effect of the PGI seems to be stronger for the Treated, and seems to moderate the slope of the MoB effect differentially for the Treated and Controls.  Indeed, if we drop the triple interaction $MoB \times PGI \times Treated$, the coefficient on the PGI becomes 0.12, though with a relatively large standard error of 0.07. Since we do not replicate this moderating effect of the PGI on MoB at later ages, we are cautious not to over-interpret this pattern.} 

Considering the $G \times E$ estimate (Treated $\times$ PGI) in Column (2), we find that the discontinuity in the entry assessment test score by treatment status is larger for those with a higher PGI: a one-standard-deviation increase in the PGI is associated with an additional 0.087 standard deviation increase in the discontinuity. This is consistent with the descriptive analysis of \autoref{fig:PGIxTreat_ea}, showing a slight divergence of the lines between the treatment and control groups. Taken at face value, the regression results suggest that genetic endowments in fact \textit{increase} the educational inequalities at age 4 that are driven by being old for grade, although the effect is only marginally significant at the 10\% level.

\begin{table}[H]
\caption{OLS estimates of the main and interaction effects of being old in the year (Treated) and the PGI for educational attainment on children's entry assessment (age 4) test scores.}
\centering
{\footnotesize
\begin{tabular}{lcccccccccccccccccccc}
\toprule
            &\multicolumn{1}{c}{(1)}         &\multicolumn{1}{c}{(2)}         \\
\midrule
Treated     &       1.138\sym{***}&       1.133\sym{***}\\
            &     (0.088)         &     (0.077)         \\
\addlinespace
PGI         &       0.155\sym{***}&       0.024         \\
            &     (0.027)         &     (0.024)         \\
\addlinespace
Treated $\times$ PGI &                     &       0.087\sym{*}  \\
            &                     &     (0.035)         \\
\addlinespace
MoB         &      -0.148\sym{**} &      -0.150\sym{**} \\
            &     (0.042)         &     (0.039)         \\
\addlinespace
Treated $\times$ MoB &       0.055         &       0.059         \\
            &     (0.045)         &     (0.045)         \\
\addlinespace
MoB $\times$ PGI     &                     &      -0.080\sym{**} \\
            &                     &     (0.021)         \\
\addlinespace
MoB $\times$ PGI $\times$ Treated&                     &       0.127\sym{***}\\
            &                     &     (0.025)         \\
\midrule
$R^2$          &       0.258         &       0.267         \\
Observations&        1094         &        1094         \\

\bottomrule
\addlinespace[.75ex]
\end{tabular}
\label{tab:MoB_ea}
}
\caption*{\footnotesize \noindent \textit{Notes:} The analysis uses a bandwidth of 3 months before and after the cutoff (i.e., June till November). Robust standard errors in parentheses, clustered by month of birth. * $p < 0.10$, ** $p < 0.05$, *** $p < 0.01$.}
\end{table}

To explore how the $G \times E$ effect changes as children age and go through the schooling system, \autoref{tab:MoB_ks} shows the analysis that uses the four Key Stage tests as the outcomes of interest. The main ``treatment effect'' is consistent with the previous literature: those who are older in their grade have test scores approximately 0.7, 0.4, 0.2 and 0.3 standard deviations higher at ages 7, 11, 14 and 16, respectively. Indeed, the treated perform better than the controls on all Key Stage tests, though the difference reduces as children age. The downward trend in test scores by month of birth is also visible in all specifications and is less steep for those born after September. Focusing on the $G \times E$ interaction term, we now find a significant \textit{negative} effect across all Key Stages other than Key Stage 3, where the coefficient is close to zero and insignificant. This finding suggests that although the treated, on average, have higher test scores, the discontinuity is smaller for those with a high PGI, reducing inequalities between the treated and control groups. 

These results are in contrast to those presented in  \autoref{tab:MoB_ea}, where the interaction term is positive and marginally significant. One potential explanation is simply that given the limited sample size for the entry assessment analysis, and the fact the influence of the PGI is still relatively limited at age 4, the estimate in \autoref{tab:MoB_ea} is imprecise ($p>0.05$). In fact, the 95\% confidence interval also spans negative values. Taken the estimates at face value, a more substantive explanation could be that the switch in the sign of the interaction term is driven by differential teacher investments in pupils. Indeed, teachers are responsible for making sure that all pupils reach a certain academic level. As August-born students are (academically) behind September-born ones, teachers may spend more time with them to bring them up to speed with others in the class.\footnote{For example, there is a clear gradient in the proportion of pupils identified as having special educational needs depending on their month of birth. The \cite{DFE2010} shows that at KS1, August-born pupils are 90\% more likely to be identified as such than their September-born peers. Pupils identified as having special educational needs generally receive more time from teachers.} The negative interaction term suggests that this additional attention given to the control group (Treated=0) may be more beneficial for those with a high PGI for educational attainment. Among individuals with a relatively higher PGI, the gap between August and September borns is smaller. This interpretation is consistent with complementarity in the production function, where higher endowments raise the productivity of subsequent investments \cite[see also][]{Muslimova2020b}. We would not find this complementarity for the entry assessment score since this test is taken \textit{before} children start school, when teachers have not yet met their pupils. If replicated in other contexts, this interpretation warrants further investigation.

\begin{table}[H]
\caption{OLS estimates of the main and interaction effects of being old in the year (Treated) and the PGI for educational attainment on children's Key Stage test scores.}
\centering
{\footnotesize
\begin{tabular}{lcccccccccccccccccccc}
\toprule
            &\multicolumn{2}{c}{Key Stage 1}            &\multicolumn{2}{c}{Key Stage 2}            &\multicolumn{2}{c}{Key Stage 3}            &\multicolumn{2}{c}{Key Stage 4}            \\\cmidrule(lr){2-3}\cmidrule(lr){4-5}\cmidrule(lr){6-7}\cmidrule(lr){8-9}
\midrule
Treated     &       0.712\sym{***}&       0.698\sym{***}&       0.392\sym{***}&       0.389\sym{***}&       0.226\sym{***}&       0.223\sym{**} &       0.285\sym{***}&       0.281\sym{***}\\
            &     (0.024)         &     (0.026)         &     (0.021)         &     (0.021)         &     (0.055)         &     (0.057)         &     (0.023)         &     (0.021)         \\
\addlinespace
PGI         &       0.261\sym{***}&       0.315\sym{***}&       0.347\sym{***}&       0.366\sym{***}&       0.358\sym{***}&       0.320\sym{***}&       0.318\sym{***}&       0.346\sym{***}\\
            &     (0.016)         &     (0.002)         &     (0.011)         &     (0.019)         &     (0.015)         &     (0.015)         &     (0.008)         &     (0.004)         \\
\addlinespace
Treated $\times$ PGI &                     &      -0.087\sym{**} &                     &      -0.052\sym{*}  &                     &       0.011         &                     &      -0.050\sym{***}\\
            &                     &     (0.026)         &                     &     (0.023)         &                     &     (0.022)         &                     &     (0.010)         \\
\addlinespace
MoB         &      -0.090\sym{***}&      -0.083\sym{***}&      -0.099\sym{***}&      -0.096\sym{***}&      -0.024         &      -0.023         &      -0.038\sym{**} &      -0.036\sym{**} \\
            &     (0.013)         &     (0.015)         &     (0.011)         &     (0.009)         &     (0.014)         &     (0.014)         &     (0.011)         &     (0.010)         \\
\addlinespace
Treated $\times$ MoB &       0.054\sym{***}&       0.049\sym{**} &       0.085\sym{***}&       0.080\sym{***}&       0.019         &       0.018         &       0.023         &       0.019         \\
            &     (0.013)         &     (0.013)         &     (0.012)         &     (0.011)         &     (0.029)         &     (0.032)         &     (0.013)         &     (0.012)         \\
\addlinespace
MoB $\times$ PGI     &                     &       0.040\sym{***}&                     &       0.011         &                     &      -0.013         &                     &       0.020\sym{***}\\
            &                     &     (0.002)         &                     &     (0.013)         &                     &     (0.012)         &                     &     (0.003)         \\
\addlinespace
MoB $\times$ PGI $\times$ Treated&                     &      -0.011         &                     &       0.015         &                     &       0.047\sym{***}&                     &      -0.002         \\
            &                     &     (0.012)         &                     &     (0.012)         &                     &     (0.011)         &                     &     (0.010)         \\
\midrule
$R^2$          &       0.175         &       0.182         &       0.147         &       0.151         &       0.147         &       0.149         &       0.142         &       0.145         \\
Observations&        3436         &        3436         &        3610         &        3610         &        3073         &        3073         &        3579         &        3579         \\

\bottomrule
\addlinespace[.75ex]
\end{tabular}
\label{tab:MoB_ks}
}
\caption*{\footnotesize \noindent \textit{Notes:} The analysis uses a bandwidth of 3 months before and after the cutoff (i.e., June till November). Robust standard errors in parentheses, clustered by month of birth. * $p < 0.10$, ** $p < 0.05$, *** $p < 0.01$.}
\end{table}

\subsection{Inference and robustness}
To check for the robustness of our results and to provide a different inference procedure, we perform a variant of \cite{Fisher1935}'s permutation test, as suggested by \cite{Buchmueller2011} and incorporated within a $G \times E$ framework in \cite{Muslimova2020b}. We run 1,000 ``placebo'' regressions where the values of both $G$ and $E$ have been randomly rearranged (permuted) across individuals. This permutation procedure breaks the connection between the $G \times E$ term and the individual outcome, and therefore, we would not expect any systematic relationship. 

\begin{figure}[H]
\centering 
\includegraphics[width=0.45\linewidth]{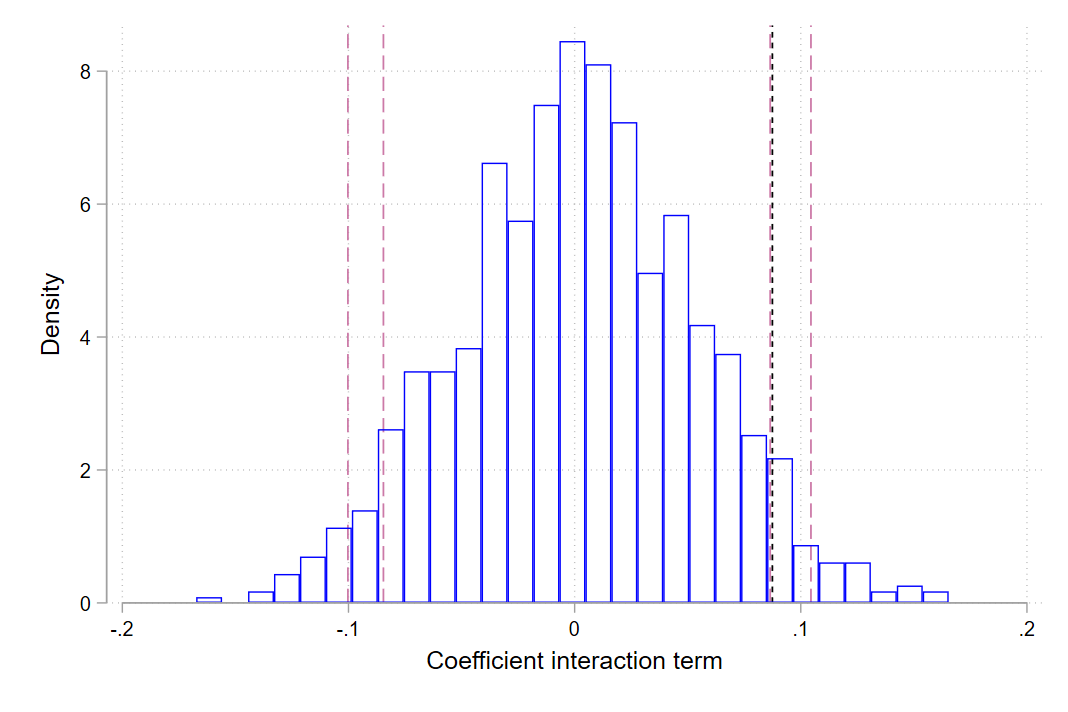}
\includegraphics[width=0.45\linewidth]{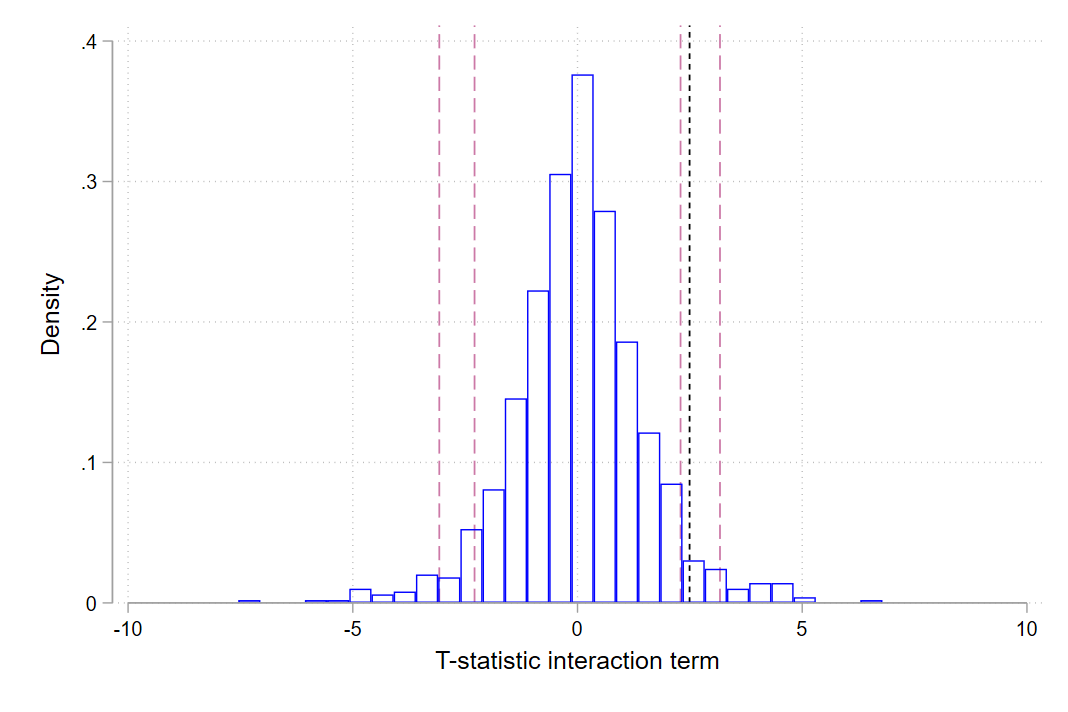} 
\caption{Distribution of the $G \times E$ coefficient and its \textit{t} statistic in 1,000 ``placebo'' regressions where $G$ and $E$ have been permuted across individuals. The dark lines represent the $G \times E$ coefficient and \textit{t} statistic from the main regression reported in \autoref{tab:MoB_ea}. The red dashed lines enclose the 90\% and 95\% probability mass of the distributions.
}
\label{fig:permutation_coef}
\end{figure}

The left-hand panel of \autoref{fig:permutation_coef} compares the coefficient for the interaction term from the main regression reported in \autoref{tab:MoB_ea} to the empirical distribution of the $G \times E$ coefficients from the ``placebo'' regressions. The right-hand panel of \autoref{fig:permutation_coef} does the same for the \textit{t} statistic. The figures show that the estimated coefficient and \textit{t} statistic lie in the upper tail of the distribution of placebo coefficients and \textit{t} statistics, respectively. They are both just outside the 90\% confidence interval, suggesting that our results are unlikely to have occurred simply by chance.

\section{Discussion and conclusion}
Recent advances in the collection and analysis of genetic data have created new opportunities for researchers to improve our understanding of how nature and nurture interact in shaping individual outcomes and thus to address some of the oldest questions in the social sciences from a new angle. In this paper, we have highlighted the exciting possibilities that the availability of genetic data provides for economics research and discussed some of the challenges that come with it in the context of studies that explore $G \times E$ interplay. With this paper, we provide practical guidance to empirical researchers, focusing on the empirical specification and the careful interpretation of models that allow $G \times E$ interplay. This contribution is very timely due to (i) reductions in the cost of genotyping, (ii) technological advances, and (iii) the public availability of PGIs for important economic traits in an increasing number of datasets. Hence, we anticipate a substantial rise in economists' use of these data over the coming years.

Having said this, there are a number of important issues to take into account in any analysis that uses PGIs and in analyses on $G \times E$ interplay. First, an estimated 79\% of participants in genetic studies are of European descent, despite this population group making up only 16\% of the global population \citep{Martin2019}. Given that the accuracy of genetic prediction decays with a larger genetic difference between the original GWAS and the prediction sample \citep{Martin2017,Scutari2016}, genetic prediction in European populations significantly outperforms that in, e.g., East Asian or African-American populations. This is driven by cross-ancestry differences in linkage disequilibrium (LD) and genetic architecture and has many implications. For $G \times E$ analysis, it implies that most new discoveries (e.g., policies that reduce the penetrance of genetic inequalities) may not be usefully extrapolated to non-Europeans. As a result, medical discoveries and social interventions based on genetic data are of greatest use to ethnic Whites. There are therefore concerns that genetic research may further widen inequalities. A concerted effort is underway to address this historical imbalance and develop new methods, such as multi-ancestry meta-analysis (MAMA) \citep{Turley2021}, that improve genetic prediction for other ancestry groups by utilizing European-ancestry GWAS results in combination with LD structures across populations. Furthermore, GWASs of non-European samples are increasingly being conducted \citep{Yengo2022}, which will also improve the predictive power of PGIs for and the generalizibility of findings for those ancestry groups.

Second, besides the overrepresentation of Europeans in genetic datasets, a second type of sample selection challenges $G \times E$ interplay studies. Many datasets with genetic information are not representative of the general population from which the participants are drawn. For example, participants of the UK Biobank are known to differ on a range of sociodemographic and lifestyle characteristics, leading to ``healthy volunteer'' bias \citep{Fry2017} and possible collider bias \citep{munafo2018collider}. Sample selection is an important issue in any empirical analysis but perhaps particularly so for genetic prediction using PGIs since this requires at least two independent samples for construction and prediction. Furthermore, since many data sources genotype their participants only in adulthood, the data implicitly incorporate survival bias. Such bias is most relevant for explorations of potentially fatal diseases such as cardiovascular disease but also affects other outcomes (such as educational attainment or social class) to the extent that mortality is correlated with such outcomes. It is important to note that such selection biases may matter less for the estimation of the environmental effect in a $G \times E$ model with exogenous $E$. Indeed, using exogenous variation in the environment via---for example---a regression discontinuity design relies on \textit{local} identification, implying that the coefficient on $E$ remains unbiased but is not necessarily generalizable to the rest of the population. 

Third, PGIs have been constructed (i) as the best \textit{linear} genetic predictor of an outcome and (ii) to maximize \textit{predictive} power. A $G \times E$ analysis is inherently \textit{non}linear and aims to investigate \textit{heterogeneity} and variability, not necessarily to maximize predictive power. Therefore, from a theoretical perspective, PGIs might not be the best proxy for $G$ in a $G \times E$ analysis. Moreover, a PGI is estimated in a first step, and therefore a `generated regressor' \citep{pagan1984econometric}. Researchers across various disciplines have started using PGIs in $G \times E$ analyses as a convenient, highly predictive and readily available proxy of genetic predispositions $G$. We follow this convention in this paper, but believe there is room for methodological advances in measuring genetic predispositions $G$ that are more robust and geared towards $G \times E$ analyses \citep[see e.g.,][]{Johnson2020.08.30.274530}. 

Fourth, and arguably most important, the potential uses of genetic data carry many societal implications and associated risks. However, simply denying the existence of genetic differences across individuals is unlikely to be the right antidote against ideas of genetic determinism or essentialism \citep{raffington2020polygenic,Harden2021}. There are ethical issues involved in working with genetic data and researchers have obligations to preserve the highest standards of privacy, confidentiality, and responsible communication. Genetic data is typically stored on secure servers, shared only with qualified academic researchers, with access to the data guarded by strict security protocols and managed by a data-use committee. PGIs, however, can be shared with fewer restrictions as it is not possible to identify individuals on their basis. Communication also matters. Researchers from social science genetics projects, such as the SSGAC, have set the current standards for responsible communication in genetics research \citep[see, e.g.,][]{lee2018gene,Ganna2019}. It is important that researchers take very seriously the need to help the public understand how to interpret research findings based on genetic data and to clarify what conclusions can and cannot be drawn from them \citep{dangerouswork}. To be clear, modern genetic analyses do \underline{not} support the notion of genetic determinism. 

There is also a reluctance by individuals to share their genetic data with governments or private companies, such as health insurers. It is important to note here that policy-makers can incorporate results from $G \times E$ analyses in formulating policies \textit{without} knowing an individual's actual genotype. In other words, for policy purposes, it is \textit{not} required that we genotype the full population \underline{nor} is it necessary to share private genetic data with the government or private companies.  Indeed, since we cannot (easily) change humans' genotype, policy-makers can only influence the environment to which individuals are exposed. Genetic data helps researchers and policy makers better understand why certain policies work or not work and for whom, allowing for the development of policies that are more robust, e.g., have desired effects for everyone. Research in economics and the social sciences on gene--environment interplay can help identify causal pathways involved in individual development and refute genetic or environmental determinism whilst identifying policy-relevant environments that can reduce socioeconomic or genetic inequalities and improve well-being in the population. 

\newpage
\bibliographystyle{apalike}
\bibliography{genes} 

\clearpage
\appendix

\section{Glossary}
\label{sec:glossary}

In this section, we provide an overview of the genetic terms and concepts used in the paper. 

\textbf{Active gene--environment correlation:} An association between genetic variation and an environment resulting from self-selection of genetically different individuals into particular environments.

\textbf{Alleles:} The nucleotides that can be present at a specific location in DNA.

\textbf{Base pairs:} Nucleotides are paired: ``A'' on one strand of the DNA always binds with ``T'' on the other strand, and ``C'' always binds with ``G''. These combinations are called base pairs.

\textbf{Candidate gene:} A polymorphism hypothesized to be associated with a particular phenotype.

\textbf{Chromosome:} A long DNA molecule. Every cell in the human body contains 23 pairs of chromosomes (22 so-called autosomal chromosomes and 1 sex chromosome).

\textbf{DNA:} Human deoxyribonucleic acid (DNA), the sequence of about 3 billion pairs of nucleotide molecules. Its double-helix structure joins two strands of DNA, where the nucleotide ``A'' binds with ``T'', and ``G'' binds with ``C''.

\textbf{Evocative gene--environment correlation:} An association between genetic variation and an environment resulting from an environmental reaction to genetic differences.

\textbf{Epigenetics:} The study of heritable phenotypic variation that does not involve changes in the DNA sequence of nucleotides.

\textbf{Gene--environment correlation:} An association between genetic variation and an environment.

\textbf{Gene:} A sequence of nucleotides in the DNA that encodes for a particular protein or proteins.

\textbf{Genetic nurture:} Parental genes influencing offspring outcomes through environmental pathways.

\textbf{Genetic variation:} Differences in the DNA among individuals; single-nucleotide polymorphisms (SNPs) constitute the most common source of genetic variation.

\textbf{Genotype:} The specific combination of base pairs (in a chromosome pair) at a particular location in the DNA sequence. If the base pairs are the same, the genotype is homozygous. If they are different, it is heterozygous.

\textbf{Genome-wide association study (GWAS):} A study in which millions of polymorphisms from the whole genome are individually tested for association with a phenotype.

\textbf{GWAS meta-analysis:} Meta-analysis of genome-wide association study (GWAS) results from different samples.

\textbf{Genome-wide significance:} The significance level at which an association is considered statistically significant in a genome-wide association study (GWAS) ($5\times10^{-8}$).

\textbf{G$\times$E interplay:} The interplay between people’s genetic makeup and the (e.g., social, biological, and economic) environment in which they live contributing to intra-individual differences.

\textbf{Heritability:} The proportion of the total variance in a phenotype that can be explained by genetic factors.

\textbf{Imputation:} Imputation of not directly genotyped genetic variation from reference panels based on linkage disequilibrium (LD).

\textbf{Linkage disequilibrium (LD):} The correlation between adjacent nucleotides in the DNA resulting from the co-inheritance of alleles.

\textbf{Locus:} A stretch of nucleotides in strong linkage disequilibrium (LD) with each other.

\textbf{Major allele:} The allele of a single-nucleotide polymorphism (SNP) that is most common in the population.

\textbf{Manhattan plot:} A plot often used to visualize genome-wide association study (GWAS) results, with genomic coordinates on the $x$ axis and the negative logarithm of the associated $p$ value for each SNP on the $y$ axis. An example is \autoref{fig:manhattan}.

\textbf{Minor allele:} The allele of a single-nucleotide polymorphism (SNP) that is least common in the population.

\textbf{Nucleotide:} The basic component molecules of DNA. Human DNA is composed of a sequence of about 3 billion pairs of nucleotide molecules. There are four different nucleotides in the DNA: adenine (A), guanine (G), cytosine (C) and thymine (T).

\textbf{Passive gene--environment correlation:} An association between genetic variation and an environment resulting from the correlation between parental genes and the environment in which the child is raised.

\textbf{Phenotype:} An observable trait of an organism.

\textbf{Pleiotropy:} The influence of one gene on two or more phenotypes.

\textbf{Polygenic indices:} The best linear genetic predictor of a phenotype, constructed as the linear combination of single-nucleotide polymorphisms (SNPs) weighted by their association with the phenotype as estimated in a genome-wide association study (GWAS).

\textbf{Polygenic trait:} A trait influenced by many genetic variants, with each having a small effect.

\textbf{Polymorphism:} Locations in the DNA where the nucleotides differ between individuals.

\textbf{Population stratification:} The presence of a systematic difference in allele frequencies between subpopulations within a population.

\textbf{Principal components:} Principal components extracted from the genetic relatedness matrix, used to control for subtle population stratification.

\textbf{Single-nucleotide polymorphism (SNP):} A single nucleotide location in the DNA that varies between individuals.

\section{Detailed discussion of biases arising from endogenous \texorpdfstring{$E$}{}} \label{appsec:bias}
\subsection{Endogeneity of \texorpdfstring{$E$}{}}
Endogeneity of $E$ may arise from four sources: reverse causality, omitted variable bias (especially correlation with parental genotype), measurement error, and correlation of the GWAS sample selection with the environment $E$. The last source of bias is discussed in the main text in \hyperref[subsec:endogenousE]{Section~\ref*{subsec:endogenousE}}. We discuss the other sources of bias here. We focus specifically on the resulting bias in the estimated coefficients of \autoref{eq:function} in a setting with gene--environment interplay. 

\paragraph{Bias arising from reverse causality.}
First, there may be reverse causality, with the outcome influencing the relevant environment. Such endogeneity will bias the estimated effect of $E$ and $G \times E$ but not the effect of predetermined $G$. 

\paragraph{Bias arising from omitted environmental variables.}
Second, omitted variable bias can arise because environments typically do not arise in isolation. For instance, education, employment, and income are correlated with each other and with unobserved confounders. Therefore, even if a significant $E$ (or $G \times E$) is found, it is difficult to identify whether the association is driven by income, employment, education, or something else  \citep{Boardman2013}. Hence, endogeneity of $E$ implies that the coefficient on $G \times E$ may in fact reflect $G \times E^*$, i.e., a causal effect of some environment $E^*$ that is correlated with $E$. This is shown in Columns (2) and (3) of \autoref{tab:Interpretation}. 

\paragraph{Bias arising from $rGE$.}
Another form of omitted variable bias, more specific to the case of gene--environment interplay, is that environments could reflect parental or one's own genes through $rGE$. Indeed, if the environment to which one is exposed is partially shaped by (parental) genes, which may also affect the outcome of interest, then it is no longer clear whether the coefficients on $E$ and $G \times E$ genuinely reflect policy relevant parameters \citep{Wagner2013rge}. This can lead to spurious detection of a $G \times E$ effect when in fact one is measuring the effect of $G \times G$ (e.g., if $E$ is shaped by one's genes through active $rGE$) or of $G \times E^*$ (through correlated environments). These scenarios are presented in the last column of \autoref{tab:Interpretation}. 

\paragraph{Bias arising from measurement error.}
A third source of endogeneity in $E$ is measurement error. If there is no $rGE$, then standard econometric theory explains how classical measurement error in $E$ will lead to attenuation bias. If there is no measurement error in $E$ but there is correlation between $G$ and $E$, the implications are more subtle: the known measurement error in PGIs as a proxy for $G$ will also lead to measurement error in $E$. Active $rGE$ implies that $G$ leads to self-selection into certain environments $E$. If in turn $G$ is measured with error, then not only is there attenuation bias in the coefficient of $G$, but also the coefficient of $E$ will be biased since the measurement error in $G$ introduces a kind of omitted variable bias in $E$.

\end{document}